\documentclass[12pt]{article}
\usepackage{setspace,graphicx,epstopdf,amsmath,amsfonts,amssymb,amsthm}
\usepackage{marginnote,datetime,enumitem,subfigure,rotating,fancyvrb}
\usepackage{hyperref,float}
\usepackage[longnamesfirst]{natbib}
\usepackage{mathtools}
\usepackage{caption}
\usepackage{float}
\usepackage{lscape}
\usepackage{graphicx}
\usepackage{flafter} 
\usepackage{tabularx} 
\usepackage{booktabs}
\usepackage{changepage}
\usepackage{setspace}
\usepackage{placeins}
\usepackage{threeparttable}
\usepackage{ragged2e}
\usepackage[export]{adjustbox}
\usepackage{stmaryrd}
\newcolumntype{Y}{>{\raggedleft\arraybackslash}X}
\usdate
\usepackage[a4paper,bindingoffset=0.2in,%
            left=0.8in,right=0.8in,top=1in,bottom=0.8in,%
            footskip=.35in]{geometry}

\usepackage{indentfirst} 
\usepackage{endnotes}    

\begin{document}

\setlist{noitemsep}  
\onehalfspacing      
\renewcommand{\footnote}{\endnote}  

\author{\large{Mykola Pinchuk}\thanks{\rm Simon Business School, University of Rochester. Email: Mykola.Pinchuk@ur.rochester.edu. \newline I would like to thank Alan Moreira, Yixin Chen, Bill Schwert, Christian Opp, Yukun Liu, Jerry Warner, Shuaiyu Chen, Pingle Wang, Xuyanda Qi, Yushan Zhuang, David Swanson and Robert Mann for helpful comments. All errors are my own.}}

\title{\bf Labor Income Risk and the Cross-Section of Expected Returns}

\date{09 January 2022}  

\maketitle
\thispagestyle{empty}

\bigskip

\normalsize
\vspace{1cm}

\centerline{\bf Abstract}

\vspace{0.5cm}

\begin{onehalfspace}  
  \noindent This paper explores asset pricing implications of unemployment risk from sectoral shifts. I proxy for this risk using cross-industry dispersion (CID), defined as a mean absolute deviation of returns of 49 industry portfolios. CID peaks during periods of accelerated sectoral reallocation and heightened uncertainty. I find that expected stock returns are related cross-sectionally to the sensitivities of returns to innovations in CID. Annualized returns of the stocks with high sensitivity to CID are 5.9\% lower than the returns of the stocks with low sensitivity. Abnormal returns with respect to the best factor model are 3.5\%, suggesting that common factors can not explain this return spread. Stocks with high sensitivity to CID are likely to be the stocks, which benefited from sectoral shifts. CID positively predicts unemployment through its long-term component, consistent with the hypothesis that CID is a proxy for unemployment risk from sectoral shifts. 
\end{onehalfspace}
\medskip

\clearpage
\setstretch{1.525}

\section{Introduction} \label{sec:Model}
The economy is a dynamic system with a permanently evolving structure. As some industries are born, other industries lose their importance. Few people in the 1970s could have predicted that within 20 years IT sector would capture the largest share of the stock market. There is no reason to expect a constant rate of this structural transformation. Naturally, accelerated structural transformation increases labor income risk for workers in underperforming industries. The COVID-19 pandemic and the resulting reallocation of labor and capital across industries is the most recent example of this transformation. This paper explores the asset pricing implications of such sectoral reallocation and related labor risk. 
\paragraph{}
The paper measures time-varying structural uncertainty using cross-industry dispersion (CID), defined as the mean absolute deviation of the returns of 49 industry portfolios. I document novel evidence that stock exposures to cross-industry dispersion (CID) predict the cross-section of expected returns. The paper finds that the firms with large return sensitivity to CID deliver smaller returns than the stocks with low sensitivity to CID. This finding suggests that in equilibrium there exists an extra demand for stocks, positively comoving with CID, implying low expected returns of these stocks. Such stocks have high momentum and investment as well as low book-to-market, consistent with the idea that they are likely to benefit from sectoral reallocation and are less risky. The long-short value-weighted portfolio, formed on the sensitivity to CID, delivers a monthly return of 49 bps with abnormal returns of 29-60 bps. The results are not explained by the other measures of volatility and uncertainty. CID subsumes a large fraction of cross-sectional return predictability of common idiosyncratic volatility CIV (Herskovic, Kelly, Lustig, Van Niewerburgh 2016). 
\paragraph{}
This evidence is consistent with several broad sources of risk, possibly related to CID. Periods of highly nonhomogeneous performance across industries (high CID) are likely to be periods of accelerated sectoral reallocation of resources and increased economic uncertainty. Times with a large wedge between winning and losing industries can proxy for periods of rapid economic transformation, when some industries become more important, while others lose relevance. High CID may indicate a temporary decrease in output and efficiency as resources are reallocated across sectors (Lilien 1982). 
\paragraph{}
More specifically, labor income risk can explain the relevance of CID to asset pricing. By construction, high CID means that some industries severely underperform the market, while others enjoy exceptionally high returns. This pattern is likely to raise a concern that the losing industry is under threat of dying. Naturally, this is a period of a negative shock to the expected lifetime labor income of the employees in underperforming industries. The fact that there are both outperforming and underperforming industries with returns sufficiently far from the market return implies an even larger labor income risk than the one during a recession. Within a recession, the common wisdom is that most of the lost jobs will be recovered at some point in the future, while a disappearance of some industry means a loss of industry-specific jobs forever. 
\paragraph{}
While many occupations are demanded across industries, they are usually low-skill occupations, less relevant to asset pricing. Losing a high-wage job constitutes a larger negative shock to expected lifetime income. Moreover, high-paid workers are likely to have more savings and produce a larger effect on the financial market. High-skill jobs (e.g., aerospace engineer) are usually confined to a few industries, implying very low across-industry labor mobility. Thus, I expect that CID captures the asset pricing effect of Cross-sectional Dispersion (CSD), consistent with labor income risk. According to the labor income risk explanation, stocks with negative covariance with CID are riskier, since they tend to fall when the risk of losing high-skill industry-immobile jobs is large.
\paragraph{}
This paper draws upon the theoretical contribution of Duffie and Constantinides (1996) to motivate CID as a state variable. Duffie and Constantinides show that in a model with heterogeneous agents, subject to uninsurable labor income shocks, the cross-sectional variance of agents' consumption growth becomes a state variable. As discussed above, the severe underperformance of their industry suggests a negative shock to the expected consumption growth of employees.
\paragraph{}
This article builds upon a body of work from macroeconomics. Lilien (1982) reports that sectoral shifts and performance dispersion result in unemployment shocks, caused by the migration of employees between firms and sectors. Loungani, Rush and Tave (1990) construct a stock market dispersion index and document that it predicts unemployment. Brainard and Cutler (1993) use the similar measure to assess the contribution of sectoral reallocation to unemployment. They find that sectoral reallocation accounts for a large share of unemployment fluctuations at long horizons. Multiple studies (Summers and Carrol 1991, Blundell, Pistaferri and Preston 2008, Guvenen and Smith 2014, Heathcote, Storesletten and Violante 2014) document that households do not completely insure their consumption from persistent shocks to the labor income. These findings highlight the importance of labor income risk for asset pricing.
\paragraph{}
The research on economic uncertainty frequently uses cross-sectional dispersion of accounting variables as a measure of uncertainty. Bloom (2009) documents a strong correlation between the indicators of macroeconomic uncertainty, including time-series measures of market volatility, the cross-sectional variance of firms` pretax profit growth and returns as well as a dispersion across macroeconomic forecasts. He argues that all of these variables proxy for financial and macroeconomic uncertainty. Sadka (2012) finds that higher earnings dispersion is associated with higher expected returns, suggesting that earnings dispersion is a state variable from an asset pricing model. Jurado, Ludvigson and Ng (2015) use cross-sectional dispersion of profit growth in order to construct the index of macroeconomic uncertainty.
\paragraph{}
This study contributes to several domains of literature. Different measures of market volatility have traditionally been used as a proxy for economic and financial uncertainty. Ang, Hodrick, Xing and Zhang (2006) empirically show that exposure of stocks to VIX is priced in the cross-section of expected stock returns. Herskovic, Kelly, Lustig, and Van Nieuwerburgh (2016) report that stocks with larger sensitivity to common idiosyncratic volatility (CIV) earn lower expected returns. They argue that CIV proxies for idiosyncratic risk faced by households and suggest labor income risk as a main channel. Eiling (2013) uses industry-specific labor income growth rates to extend human capital CAPM (Jagannathan and Wang 1996) and shows that her empirical model can price 25 size and book-to-market portfolios. Eiling, Kan, and Sharifkhani (2021) use production-based asset pricing model to motivate the predictability of market returns by sectoral labor reallocation shocks. 
\paragraph{}
This paper is closely related to the literature on cross-sectional dispersion (CSD), defined as a standard (mean absolute) deviation of the cross-section of stock returns. Bekaert and Harvey (1997, 2000) use CSD as a measure of the development and maturity of the stock market. Stivers (2003, 2006, 2010) finds that CSD can predict idiosyncratic volatility as well as returns of value and momentum premia. Maio (2013, 2016) reports that the cross-sectional standard deviation of returns of the portfolios sorted on size and book-to-market can predict market returns. The closest paper to this one is Verousis and Voukelatos (2018). They find that the stocks with high sensitivity to CSD offer lower returns in their 1996-2012 sample. Due to the more idiosyncratic nature of CSD, the authors explain the findings by idiosyncratic risk without elaborating on or providing evidence for their explanation.
\paragraph{}
This paper uses 1963-2018 sample and shows that high sensitivities to CID predict lower expected returns. I argue that this predictability stems from return dispersion across industries, proxying for labor income risk due to sectoral shifts. These results suggest that only the across-industry component of CSD (i.e., CID) is priced. The long-short portfolio, formed on CID, controlling for CSD, delivers 29 bps (t-statistic=2.1) monthly abnormal returns, while the long-short portfolio, formed on CSD, controlling for CID, produces 10 bps (t-statistic=0.5) returns. Therefore, these results suggest that the cross-sectional predictability of returns by CSD is not driven by idiosyncratic risk, relevant for underdiverisified investors. CSD is a manifestation of a more fundamental economic force - labor income risk from structural shifts, proxied by CID. 
\paragraph{}
Consistent with this explanation, high CID predicts high aggregate unemployment at the quarterly frequency. The results suggest that the predictive power of CID for unemployment growth is driven by the long-term component of unemployment growth. Unlike aggregate demand shocks, sectoral reallocation shocks are likely to have a permanent effect on unemployment growth (Brainard and Cutler 1993). Consistent with CID, proxying for labor income risk from sectoral shifts, CID strongly predicts long-term unemployment but has little relation with its short-term component.  Overall, these results support the explanation of CID premium by industry-specific human capital risk from sectoral shifts.


\section{Data and Dispersion Measures} \label{sec:Model}

\subsection{Sample}

The paper uses stock price data from CRSP and accounting data from Compustat. The sample encompasses 1963-2019. I download macroeconomic data from St. Louis Federal Reserve Bank. Bureau of Labor Statistics (BLS) is the source of unemployment data.
I consider common stocks traded at NYSE, NASDAQ or AMEX and perform the main asset pricing tests at the monthly frequency. The sample of stock returns consists of 3058110 firm-month observations with nonmissing returns. In order to exclude small and illiquid stocks from the analysis, I restrict the sample to the stocks with price above \$5 at the end of the previous year and the stocks with market capitalization above \$50 million in 2018 dollars. After this restriction the sample consists of 1841271 firm-month observations. All the analysis below uses excess returns.
\paragraph{}
Kenneth French`s website is a source of data on stock market factors as well as industry classifications and returns of industry portfolios. I download uncertainty indices, constructed by Jurado, Ludvigson and Ng (2015) from Sydney Ludvigson`s website. 

\subsection{Construction of Dispersion Measures}

The paper defines cross-industry dispersion as the mean absolute deviation of returns across 49 industry portfolios at any given period.
\begin{equation}
CID_t = \frac{1}{N}\sum^{N}_{i=1}{|R_{it}-R_{MKT,t}|}.
\end{equation}
I use monthly value-weighted returns of Fama-French 49 industry portfolios. I compute CID every month during the period, spanning 1927-2020. Since some industries have few firms at the beginning of the sample, the paper computes CID across the industries with at least 10 firms. I proxy for the market return with the value-weighted market return across all firms from CRSP (vwretd). I use the similar approach to calculate Cross-Sectional Dispersion (CSD) and Within-Industry Dispersion (WID).
\paragraph{}
The Figure 1 plots the time series of CID. To be consistent with asset pricing literature, I focus on CID after 1963. The time series of CID exhibits two large secular increases. The first long-term spike took place within 1998-2003 and is likely associated with the dotcom boom and the following decline. The second spike was recorded during 2007-2009, coinciding with the recent financial crisis. Both events are consistent with the hypothesis that CID proxies for some dimension of uncertainty and is related to the rate of sectoral shifts in the economy. Intuitively, CSD is a sum of CID and WID, so Figure 1 facilitates comparison between these variables. Notice large spike in CID during March-April 2020. This jump in CID reflects heterogeneous impact of COVID-19 pandemic and related mitigation measures on different industries. 
\paragraph{}
The next step of the analysis will be to estimate sensitivities of the stocks to fluctuations in CID. To facilitate the interpretation of CID as a state variable from consumption-based asset pricing model and avoid econometric issues, I difference CID, using the same specification as in Pastor and Stambaugh (2003). That is, I calculate first difference of CID and estimate its residuals from AR(1) model:
\begin{equation}
\Delta(CID_t) = \gamma_0 + \gamma_1 \Delta(CID_{t-1}) + \gamma_2 CID_{t-1} + u_t.
\end{equation}
Abusing notation, I refer to the residual $\hat{u}_t$ as $CID_t$ in the rest of the paper. 1-lag autocorrelation of CID is -0.05, implying very low persistence. Thus, I can use $CID_t$ as an explanatory variable in predictive regressions without spurious regression issues (Novy-Marx 2013). 
\paragraph{}
Table 1 reports correlations of CID with other measures of uncertainty and volatility. Since all time series are differenced according to Pastor and Stambaugh (2003) specification, the correlations are relatively low. CID has positive correlation with all the variables, ranging from 0.07 to 0.30. The results suggest that CID is distinct from, though positively related to frequently used uncertainty and volatility measures.

\section{Asset Pricing Results} \label{sec:Model}

This section documents and discusses the main results of the paper. First, I explain how I estimate the sensitivity of stock returns to CID. Then, I form decile portfolios on $\beta_{CID}$ and compute their returns. Finally, the section discusses the asset pricing performance of different measures of cross-sectional dispersion.

\subsection{Estimation of $\beta_{CID}$}
I use monthly stocks returns in order to obtain estimates of their sensitivities to CID. While using daily returns produces more precise estimates, it hampers interpretation of CID as labor income risk. The paper uses two years of monthly excess returns to estimate betas from the following time series regression:
\begin{equation}
R_{it}=\alpha+\beta CID_t + \epsilon_t.
\end{equation}
Since controlling for market returns has little effect on the results, I use simpler specification (3). I estimate betas over the period 1963-2018. To compensate for the extreme values of $\hat{\beta}_{CID}$, possibly resulting from imprecise estimation, I winsorize $\hat{\beta}_{CID}$ at 1\% and 99\% percentiles. Section 5.3 verifies that the results are not due to factors, omitted during estimation of $\beta_{CID}$.  

\subsection{Portfolios, formed on $\beta_{CID}$}

Every month, I sort the stocks into quintile portfolios based on their $\beta_{CID}$. I update estimates of $\beta_{CID}$ and rebalance portfolios every month. 
Table 2 reports mean values of characteristics of decile $\beta_{CID}$-sorted value-weighted portfolios. Stocks in the highest quintile are somewhat larger, more liquid, have faster asset growth and higher past returns. Furthermore, high $\beta_{CID}$ stocks have high momentum and investment as well as low book-to-market. These results are consistent with the idea that $\beta_{CID}$ stocks are likely to have benefited from sectoral shifts in the past. Since there are persistent long-term trends in the sectoral composition of the economy, it is reasonable to assume that the stocks, which benefited from sectoral shifts in the recent past will continue benefiting from sectoral reallocation in the near future. In unreported results, I show that high $\beta_{CID}$ stocks have very high cumulative past returns over any time period up to 10 years. This observation is consistent with those stocks benefiting from long-term trends in sectoral composition of the economy.
\paragraph{}
Table 3 describes excess returns of value-weighted portfolios, formed on $\beta_{CID}$. Returns of value-weighted portfolios, sorted on $\beta_{CID}$, decrease monotonically. The long-short value-weighted portfolio generates 49 bps average monthly returns with t-statistic of 3.19. 
\paragraph{}
Table 4 presents abnormal returns of the value-weighted long-short quintile portfolio with respect to most frequently used factor models. Controlling for market beta does not have significant effect on the CID premium. T-statistic of these abnormal returns stays comfortably below -3, meeting the challenge of Harvey et al. (2016). Fama-French 5 factor model, augmented with momentum and short-term reversal factors, leaves abnormal returns of 50 bps (t-statistics = -3.26). While the loadings on size, vale and profitability factors can explain a fraction of returns of this trading strategy, the loadings on market and momentum factors make these returns even more anomalous (Table 5). Profitability is the only factor, which can explain large part of CID premium. The results are consistent with the idea that CID premium is the price of some risk, unexplained by commonly used factor models. The next section empirically shows that high CID is associated with larger unemployment risk for workers with industry-specific human capital. High $\beta_{CID}$ stocks are natural hedges against such labor risk from sectoral shifts.
\paragraph{}
Table 6 reports the returns of long-short portfolios, double-sorted on size and CID. Even controlling for size, the long-short portfolio from tercile sorts on CID delivers statistically significant abnormal returns of 17-40 bps. Somewhat weaker results are likely driven by both size effect and a low precision of $\beta_{CID}$ estimates, combined with very coarse sorts.
\paragraph{}
Monthly frequency, used to estimate $\beta_{CID}$, raises the question of the trade-off between economic intuition and statistical precision. While labor risk story is more intuitive at monthly CID frequency, the estimates of sensitivity of returns to other variables at monthly frequency are notoriously imprecise. In unreported results, $\beta_{CID}$, estimated at daily frequency, produces larger return spreads and strong positive relationship between pre-ranking betas and post-ranking betas.

\subsection{Cross-sectional Results}

To check a robustness of the cross-sectional predictability of returns by $\beta_{CID}$, I employ Fama-MacBeth regressions. Table 7 reports the results of Fama-MacBeth regressions across decile $\beta_{CID}$-sorted portfolios. Controlling for stock characteristics, $\beta_{CID}$ is a significant negative determinant of excess returns. Its coefficient varies between -0.08 and -0.12, corresponding to 1.0\% to 1.5\% smaller annualized returns. Since the standard deviation of $\beta_{CID}$ is 3.22, the results suggest that 1 standard deviation increase in $\beta_{CID}$ is associated with 3.2\%-4.6\% smaller annualized returns. These numbers suggest economically significant price of risk, associated with CID. These estimates of the price of risk are very similar to the estimates from nonparametric analysis using time-series regressions.

\subsection{Performance of other dispersion measures}

Verousis and Voukelatos (2018) find that in 1996-2012 sample lower CSD is associated with high excess returns. The obvious question is whether CSD premium is driven by its across-industry component (CID) or within-industry component (WID). To answer this question, I perform double sorts on WID and CID. Table 8 reports the results with value-weighted portfolios from 5x5 double sorts. I construct the long-short portfolio from the sort on CID, controlling for WID, and vice versa. Except the case of CAPM, the long-short portfolio from WID sorts delivers abnormal returns between 21 and -15 bps, not significantly different from zero. The long-short portfolio from CID sorts produces abnormal returns between 30 and 14 bps, which are  significant in three cases. These results imply that CID is better than WID in capturing a priced dimension of cross-sectional dispersion. 
\paragraph{}
Table 9 reports the similar results from 5x5 double sorts on $\beta_{CID}$ and $\beta_{CSD}$. The long-short portfolio from CSD sorts produces negative abnormal returns with respect to Fama-French 5 factor model, augmented with momentum factors. Since these returns have the sign, inconsistent with the idiosyncratic risk explanation of Verousis and Voukelatos, the results clearly suggest that CSD predicts the cross section of expected returns only through CID. Controlling for CSD, CID long-short portfolio generates abnormal returns between 12 and 30 bps. Overall, the results suggest that CID reflects a priced component of cross-sectional volatility.

\subsection{CID and idiosyncratic volatility}
As Table 1 suggest, sensitivity of stocks to CID is likely to be strongly correlated with the sensitivity to other volatility measures, such as market volatility and CIV (Herskovic, Kelly, Lustig and Van Nieuwerburgh 2016). Since Herskovic et al. (2016) argue that CIV captures household labor risk, it is important to compare asset pricing relevance of CID and CIV. 
\paragraph{}
Table 11 reports the results of double-sorts on $\beta_{CID}$ and $\beta_{CIV}$. 5x5 independent double sort allows to form long-short portfolios on $\beta_{CID}$, controlling for $\beta_{CIV}$ and vice versa. While long-short portfolio, formed on $\beta_{CID}$ generates economically and statistically significant return of 40 bps, the long-short portfolio, formed on $\beta_{CIV}$, has only 8 bps returns and fails to consistently produce abnormal returns. Its abnormal returns range between -5 and 5 bps and are not statistically significant. On the other hand, the long-short portfolio, formed on $\beta_{CID}$, delivers abnormal returns between 20 and 54 bps with t-statistics above 2 with respect to most factor models. Overall, the results suggest that CID mostly subsumes abnormal component of CIV premium.
\paragraph{}
In order to obtain further evidence on asset pricing performance of CID and CIV, I employ spanning tests. Tables 12 and 13 document their results using CID and CIV factors, constructed as the long-short quintile portfolios from univariate sorts on these variables. Controlling for CIV factor decreases abnormal returns of CID factor from 50 to 47 bps with t-statistic of 3. On the other hand, CID factor weakens performance of CIV factor from 15 to 5 bps abnormal returns (t-statistic of 0.3). 
\paragraph{}
CID can explain significant fraction of CIV premium, while the results from spanning tests suggest that CID factor mostly subsumes CIV factor. Overall, CID and CIV appear to share significant common component, which is better explained by CID. The remaining asset pricing power of these two measures seems to come from different sources. As a measure of time-series volatility, CIV additionally proxies for some form of time-varying uncertainty, while CID measures cross-section dispersion and is likely to proxy for the rate of sectoral transformation of the economy.

\section{Economic Channel: Labor Income Risk} \label{sec:Model}

\subsection{Economic Mechanism}

Constantinides and Duffie (1996) develop the model with agents, subject to heterogeneous uninsurable income shocks. They show that the cross-sectional variance of individual consumption growth enters Euler equation. While the model analytically proves the relevance of the cross-sectional variance of consumption growth for asset pricing, it remains silent on its economic channels and meaning. This story naturally fits macrolabor literature which uses different measures of dispersion of stock returns to proxy for unemployment shocks, resulting from structural shifts.
\paragraph{}
Fluctuations in individual labor income can arise from two channels: a change in wage and a change in employment. Larger magnitude of labor income shocks from employment loss as well as downward wage rigidity suggests that unemployment risk is the most important component of labor income fluctuations for asset pricing. Macroeconomic literature views the changes in aggregate employment as the result of two forces: aggregate shocks and reallocation shocks. Lilien (1982) introduces the sectoral shifts hypothesis entailing that higher magnitude of sectoral shifts leads to higher unemployment by increasing amount of labor reallocation. While Lilien uses the variance of employment growth across sectors to measure the reallocation, Loungani, Rush and Tave (1990) argue that the variance of stock returns across sectors is more precise and forward-looking measure. 
\paragraph{}
This paper uses CID to capture a divergence in performance across industries, indicative of a change in sectoral composition of the economy. Therefore, periods of high CID are indicators of increased risk of losing lobs in underperforming industries. While many occupations are not industry-specific (e.g., janitor), the others are closely related to few industries (e.g., aerospace engineer). A potential disappearance of underperforming industry constitutes large labor income risk for such high-skill industry-specific occupations. Addison and Portugal (1989) as well as Jacobson, LaLonde and Sullivan (1993) document that workers experience large wage decrease after switching industry. 
\paragraph{}
According to this story, changes in CID should have positive relationship with both contemporaneous and future unemployment growth. Since a positive shock to CID reflects an increase in the rate of sectoral reallocation of resources and labor, it should be associated with layoffs. Therefore, I expect a positive relationship between CID changes and unemployment changes. Since stock prices have more forward-looking nature and move faster than measures of real economic activity, the firm's employment decisions will respond to stock returns with some lag. Thus, we should expect positive predictability of unemployment growth by CID changes.

\subsection{Results}

Table 13 shows that CID predicts unemployment at the quarterly frequency, controlling for other stock market-based predictors of business cycles.
Using the sample of 1948-2019, I regress unemployment growth on the changes in CID in the previous quarter. The first two columns report positive and statistically significant coefficients (t-statistic of 2.77 and 2.59), suggesting that the results are robust to known stock market predictors of business cycle. These results are consistent with the findings of Loungani, Rush, Tave (1990) and Brainard, Cutler (1993) and suggest that CID can forecast unemployment growth. 
\paragraph{}
Labor risk explanation of CID premium is based on the idea that CID is a proxy for more permanent component of unemployment risk, arising from sectoral shifts as opposed to aggregate shocks. Macroeconomic literature (Lilien 1982, Loungani et al. 1990) argues that while unemployment from aggregate shocks have transitory nature, sectoral shifts have more long-term effects on unemployment. For example, if small short-term aggregate shock hits the economy, then struggling firms in different industries are likely to fire workers to cut the costs. As long as there are no industries, disproportionally affected by this shock, it will not cause sectoral reallocation of labor. Successful firms are likely to hire workers, fired from struggling firms in the same industries. Therefore, such short-term aggregate shocks are unlikely to have a major effect on long-term unemployment.
\paragraph{}
The last four columns of Table 13 show that the predictability of unemployment growth by CID changes is mostly driven by long-term component of unemployment growth. CID predicts growth in long-term unemployment (t-statistics 2.62), but can not predict growth in short-term unemployment. Adjusted $R^2$ of these regressions provide even stronger results. While CID explains 4\% of the variation in future long-term unemployment growth, adding aggregate market predictors (first and second moments of market returns) increases this fraction only to 6\%. On the other hand, CID has no explanatory power for the variation in future short-term unemployment growth as evidenced by zero adjusted $R^2$. However, adding aggregate market controls increases the fraction of explained variation to 17\%. Tables 20-22 show that these results are robust to controlling for more measures of uncertainty. With more controls predictability of short-term unemployment goes to zero, while predictability of long-term unemployment becomes stronger and coefficient on CID reaches 3.2 with t-statistic of 3.1.
\paragraph{}
These results suggest that CID is more important predictor of long-term unemployment than aggregate stock market variables are, while these aggregate variables are much better predictors of short-term unemployment growth. These findings are consistent with CID reflecting sectoral shifts and aggregate stock market variables accounting for aggregate short-term shocks. Thus, the evidence supports the explanation of CID premium by long-term unemployment risk from sectoral shifts.

\section{Further results} \label{sec:Model}
\subsection{Robustness to other industry classifications}

Since Fama-French (FF) 49 industry classification is arbitrary, I explore the robustness of the findings to industry definitions at the varying levels of coarseness. I compute CID using FF30, FF17, FF10 and FF5 industry definitions. Figure 3 reports CID premium at the different levels of industry aggregation. The results change a little as industry coarseness increases. Surprisingly, we observe large CID premium even as the paper computes CID across 5 industry portfolios. The results suggest that CID premium reflects the variation of returns between few broad industries. 
\paragraph{}
Figure 3 reports that the results are robust to finer industry classifications, such as the first 2, 3, 4 digits of company SIC code. Since in such fine classifications often there are very few firms in a particular industry group, I compute CID using only industries, containing at least 5 firms. The median number of such industries are 63, 174 and 212 for SIC 2, SIC 3 and SIC 4 classifications respectively. The Figure 3 shows that mean returns of $\beta_{CID}$-sorted quintile long-short portfolios continue increasing as we use finer industry classifications. This is consistent with table 8, indicating small return spread in WID dimension using FF49 industries. As we refine the industry sorts, this variation gets reflected into CID. However, consistent with the fact that WID dimension from FF49 industries does not generate large abnormal returns, using very fine industry classifications to estimate CID does not increase abnormal returns.
\paragraph{}
Table 14 reports the returns of long-short portfolios, double-sorted on CID and WID using FF5 industry classification. Now CID dimension is associated with smaller return spread, ranging between -0.01 and 0.27. This spread is statistically significant only with respect to two factor models. These results are consistent with our intuition and show that the use of very coarse industry classification leads to a loss of information, contained in CID. Since dividing firms into 5 sectors leaves relatively heterogeneous firms within each sector, a part of the variation, previously attributed to CID, is now reflected in WID. These findings are different from Table 8, which reports that CID from FF49 classification is much more relevant for asset pricing than WID. 

\subsection{Robustness to other uncertainty measures}

Table 1 reports the correlations between CID and other measures of uncertainty: VIX, market volatility, financial uncertainty (Jurado, Ludvigson and Ng 2015), macroeconomic uncertainty (Jurado, Ludvigson and Ng 2015) and common idiosyncratic volatility (Herskovic, Kelly, Lustig and Van Nieuwerburgh 2016). Consistent with the hypothesis that all these variables capture financial uncertainty, CID has a positive correlations with all of them. However, the largest correlation of CID with these variables is 30\%, implying that CID is not just a noisy proxy for these uncertainty measures.
\paragraph{}
Table 10 reports the relationship between CID premium and aforementioned uncertainty variables. I use 5x5 double sorts to control for sensitivities to these uncertainty measures. Double sorts on $\beta_{CID}$ and $\beta_{Vol}$ indicate that CID is associated with large risk premium, robust to controlling for asset pricing factors. The long-short portfolio, constructed on $\beta_{CID}$ sorts, produces abnormal returns between 34 and 60 bps and t-statistics above 2.4. While sorts on $\beta_{Vol}$ produce return spread, it does not seem robust to different factor models. Panels B and C show that $\beta_{CID}$ subsumes the relationship between stock sensitivities to financial and macroeconomic uncertainty (Ludvigson 2015) and expected returns. After controlling for $\beta_{CID}$, these uncertainty measures fail to generate return spread of correct sign. On the other hand, in 11 out of 12 specifications, CID spread is highly significant, ranging between 28 and 67 bps.
$\beta_{VIX}$ is the only variable, controlling for which eliminates statistical significance of the CID premium. Controlling for VIX exposure, long-short portfolios, formed on $\beta_{CID}$, deliver abnormal returns between -11 and +28 bps. 

\subsection{Common factors in industry portfolios}

One possible explanation of documented results is the difference in loadings on common factors across industries. According to this hypothesis, the time-series variation in CID merely reflects the variation in asset pricing factors. For example, if there is a large heterogeneity in loadings on the profitability factor across industries, then periods of very large or very low returns of the RMW factor will mechanically generate large CID.
\paragraph{}
To test this explanation, I use abnormal returns of industry portfolios with respect to Fama-French 5 factor model to calculate CID. Tables 15 reports that hedging factor exposures has no effect on quintile returns. Similarly, Table 16 suggests that controlling for factor loadings across industries has small effect on a magnitude of abnormal returns. Abnormal returns range between 22 and 57 bps.

\subsection{$\beta_{CID}$ at daily, monthly and quarterly frequency}
Labor risk explanation of CID premium crucially depends on our ability to ex ante identify stocks, which hedge unemployment shocks, driven by sectoral shifts. In other words, the necessary condition for this explanation is a positive relationship between pre-ranking and post-ranking $\beta_{CID}$. 
\paragraph{}
Decades of asset pricing research show that it is very hard to obtain accurate estimates of covariances between individual stocks and a proxy for some state variable at monthly frequency. Estimating betas at daily frequency is increasingly common. Thus this subsection explores covariance between CID and stock returns at different frequencies. 
\paragraph{}
According to labor risk explanation, investors exhibit hedging demand for the stocks, which tend to appreciate in value during periods of accelerated sectoral reallocation of labor and capital. Therefore, this story is the most consistent with monthly or quarterly $\beta_{CID}$ being priced in the cross-section of expected stock returns. Figure 4 reports the relationship between pre-ranking and post-ranking $\beta_{CID}$, estimated at monthly frequency. I estimate pre-ranking betas using rolling 24-months window and use full sample to estimate post-ranking betas of quintile portfolios. Post-ranking betas seem to be flat. Using different window lengths does not achieve positive relationship between pre- and post-ranking betas.
\paragraph{}
Tables 24 and 25 show the returns of quintile portfolios, formed on $\beta_{CID}$, estimated at quarterly frequency. While returns of long/short portfolio decrease in magnitude, alphas with respect to 3 out of 6 factor models remain statistically significant at 5\% significance level. Figure 5 shows slightly negative relationship between pre-ranking and post-ranking betas, estimated at quarterly frequency.
\paragraph{}
Alternatively, I used 2 years of the recent daily returns to estimate $\beta_{CID}$. Table 26 documents that returns of long-short portfolio remain significant at 31 bps per month. Abnormal returns of this trading strategy are significant with respect to 4 models out of 6. Most importantly, Figure 6 shows that pre-ranking $\beta_{CID}$ perfectly predicts post-ranking $\beta_{CID}$, when estimated at daily frequency. 
\paragraph{}
Table 28 explores possibility to use daily $\beta_{CID}$ as more accurate proxy for monthly $\beta_{CID}$. Surprisingly, long-short portfolios, built using the two betas, exhibit slightly negative correlation. Since it is unlikely that unemployment risk from sectoral shifts varies a lot at daily frequency, it is possible that the two trading strategies are driven by different economic forces. Uncertainty is one of the possible explanation of CID premium, when measured at daily frequency.
\paragraph{}
Lack of positive relationship between pre-ranking and post-ranking $\beta_{CID}$ poses a challenge to rational explanation of CID premium by unemployment risk. One possible explanation is investors` errors in predicting $\beta_{CID}$. This story, in which investors require premium for holding the stocks which appear though are not risky, violates rational expectations. 

\subsection{Variation in employment across industries and CID}

According to the explanation of CID premium by unemployment risk from sectoral shifts, periods of increased CID precede times of accelerated sectoral reallocation. We can use employment across industries to measure such sectoral shifts.
\paragraph{}
Bureau of Labor Statistics (BLS) conducts Current Population Survey (CPS) at monthly frequency. CPS contains information on employment across industries, defined by their NAICS codes. Before 1989 CPS has very coarse sectoral classification with 16 industries. After 1990 CPS contains employment data at much finer level of NAICS classification, allowing to match them with returns of FF49 industries. \paragraph{}
Under labor risk hypothesis, returns of industry portfolios are leading indicators of employment within these industries. Therefore, risk of sectoral shifts increases when returns across industry portfolios start diverging and not necessarily when workers lose jobs in underperforming industries.
\paragraph{}
We can test this hypothesis by using returns of industry portfolios to predict employment growth in those industries. The following two paragraphs will discuss predictability of emloyment growth in 14 industries by their returns. Since CPS is based on NAICS industry classification and this classification i very inconvenient to import, for now I precisely matched 14 industries from CPS to respective FF49 industries.
\paragraph{}
Figure 7 reports t-statistics of coefficients on lagged industry returns in monthly-frequency predictive panel regressions. Both employment growth and return of each industry are in excess of respective aggregate variables. While there is no contemporaneous relation between return and employment growth, there is significant predictability using returns for up to 3 lags. This is consistent with economic intuition: industries, underperforming in the current month, do not start firing employees immediately. Industry returns seem to lead industry employment growth by 2-3 months. This result is consistent with CID predicting unemployment at quarterly frequency. It appears that quarterly frequency is the optimal frequency to capture fluctuations in unemployment risk due to sectoral shifts.
\paragraph{}
Table 29 documents results of predictive panel regressions at quarterly frequency. The first columns shows that in 1990-2019 sample, industry return over the last quarter is significant predictor of employment growth in the current quarter. The next two columns report the results in subsamples with quarters, where industry returns are negative(positive). Most of this predictability comes from the subsample with negative stock returns. This result implies that while underperforming industries significantly cut labor force, winner industries increase number of employees to smaller degree. Therefore, industry returns contain more information about firing rather than hiring decisions of firms within industry.

\vspace{1cm}

\section{Conclusion} \label{sec:Model}

This paper introduces CID as a measure of labor income risk from sectoral shifts and explores its asset pricing implications. Stocks with low $\beta_{CID}$ are more risky since they decline precisely when investors are subject to increased labor income risk. These stocks have low momentum and investment as well as high book-to-market and are likely to suffer from sectoral shifts. Consistent with this idea, these stocks produce higher returns. The value-weighted long-short portfolio, formed on $\beta_{CID}$, delivers 49 bps monthly returns. Abnormal returns range between 29 and 63 bps. A heterogeneity in factor loadings across industry portfolios does not explain these findings. The results are not explained by known measures of uncertainty and labor income risk, such as volatility and VIX (Ang, Hodrick, Xing and Zhang 2006), macroeconomic and financial uncertainty (Jurado, Ludvigson and Ng 2015), or common idiosyncratic volatility (Herskovic, Kelly, Lustig and Van Nieuwerburgh 2016). I show that the previously documented cross-sectional predictability of returns by cross-sectional dispersion (CSD, Verousis and Voukelatos 2018) is a manifestation of the CID premium. Controlling for CID, CSD loses its asset pricing relevance, implying that CID reflects a priced component of CSD.
\paragraph{}
These findings are consistent with the hypothesis that CID premium arises due to labor income risk from sectoral shifts. This paper uses CID as a proxy for a rate of sectoral shifts, leading to unemployment risk in underperforming industries. Consistent with macroeconomic literature, I find that CID predicts unemployment. This predictability is mostly driven by medium-term and long-term unemployment, in line with the more permanent effect of sectoral reallocation shocks on unemployment. These results support the idea that CID premium is driven by industry-specific human capital risk due to sectoral shifts.

\newpage
\section{References:}
\begin{enumerate}

    \item{Addison, John T., and Pedro Portugal. "Job displacement, relative wage changes, and duration of unemployment." Journal of Labor economics 7, no. 3 (1989): 281-302.}
    \item{Ang, Andrew, Robert J. Hodrick, Yuhang Xing, and Xiaoyan Zhang. "The cross‐section of volatility and expected returns." The Journal of Finance 61, no. 1 (2006): 259-299.}
    \item{Becker, Gary S. "Investment in human capital: A theoretical analysis." Journal of political economy 70, no. 5, Part 2 (1962): 9-49.}
    \item{Bekaert, Geert, and Campbell R. Harvey. "Emerging equity market volatility." Journal of Financial economics 43, no. 1 (1997): 29-77.}
    \item{Bekaert, Geert, and Campbell R. Harvey. "Foreign speculators and emerging equity markets." The Journal of Finance 55, no. 2 (2000): 565-613.}
    \item{Bloom, Nicholas. "The impact of uncertainty shocks." econometrica 77, no. 3 (2009): 623-685.}
    \item{Blundell, Richard, Luigi Pistaferri, and Ian Preston. "Consumption inequality and partial insurance." American Economic Review 98, no. 5 (2008): 1887-1921.}
    \item{Brainard, S. Lael, and David M. Cutler. "Sectoral shifts and cyclical unemployment reconsidered." The Quarterly Journal of Economics 108, no. 1 (1993): 219-243.}
    \item{Carroll, Christopher D., and Lawrence H. Summers. "Consumption growth parallels income growth: some new evidence." In National saving and economic performance, pp. 305-348. University of Chicago Press, 1991.}
    \item{Connolly, Robert, and Chris Stivers. "Momentum and reversals in equity‐index returns during periods of abnormal turnover and return dispersion." The Journal of Finance 58, no. 4 (2003): 1521-1556.}
    \item{Connolly, Robert, and Chris Stivers. "Information content and other characteristics of the daily cross-sectional dispersion in stock returns." Journal of Empirical Finance 13, no. 1 (2006): 79-112.}
    \item{Constantinides, George M., and Darrell Duffie. "Asset pricing with heterogeneous consumers." Journal of Political economy 104, no. 2 (1996): 219-240.}
    \item{Eiling, Esther. "Industry‐specific human capital, idiosyncratic risk, and the cross‐section of expected stock returns." The Journal of Finance 68, no. 1 (2013): 43-84.}
    \item{Eiling, Esther, Raymond Kan, and Ali Sharifkhani. "Sectoral labor reallocation and return predictability." Rotman School of Management Working Paper 2602215 (2021).}
    \item {Fama, Eugene F., and James D. MacBeth. "Risk, return, and equilibrium: Empirical tests." Journal of political economy 81, no. 3 (1973): 607-636.}
    \item {Fama, E. F., \& French, K. R. (2015). A five-factor asset pricing model. Journal of Financial Economics, 116(1), 1–22. }
    \item{Goyal, Amit, and Pedro Santa‐Clara. "Idiosyncratic risk matters!." The Journal of Finance 58, no. 3 (2003): 975-1007.}
    \item{Guvenen, Fatih, and Anthony A. Smith. "Inferring labor income risk and partial insurance from economic choices." Econometrica 82, no. 6 (2014): 2085-2129.}
    \item {Harvey, Campbell R., Yan Liu, and Heqing Zhu. "… and the cross-section of expected returns." The Review of Financial Studies 29, no. 1 (2016): 5-68.}
    \item{Heathcote, Jonathan, Kjetil Storesletten, and Giovanni L. Violante. "Consumption and labor supply with partial insurance: An analytical framework." American Economic Review 104, no. 7 (2014): 2075-2126.}
    \item{Herskovic, Bernard, Bryan Kelly, Hanno Lustig, and Stijn Van Nieuwerburgh. "The common factor in idiosyncratic volatility: Quantitative asset pricing implications." Journal of Financial Economics 119, no. 2 (2016): 249-283.}
    \item{Hoberg, Gerard, and Gordon Phillips. "Text-based network industries and endogenous product differentiation." Journal of Political Economy 124, no. 5 (2016): 1423-1465.}
    \item{Jagannathan, Ravi, and Zhenyu Wang. "The conditional CAPM and the cross‐section of expected returns." The Journal of finance 51, no. 1 (1996): 3-53.}
    \item{Jacobson, Louis S., Robert J. LaLonde, and Daniel G. Sullivan. "Earnings losses of displaced workers." The American economic review (1993): 685-709.}
    \item {Jensen, Michael C., Fischer Black, and Myron S. Scholes. "The capital asset pricing model: Some empirical tests." (1972).}
    \item{Jorgensen, Bjorn, Jing Li, and Gil Sadka. "Earnings dispersion and aggregate stock returns." Journal of Accounting and Economics 53, no. 1-2 (2012): 1-20.}
    \item{Jurado, Kyle, Sydney C. Ludvigson, and Serena Ng. "Measuring uncertainty." American Economic Review 105, no. 3 (2015): 1177-1216.}
    \item{Lilien, David M. "Sectoral shifts and cyclical unemployment." Journal of political economy 90, no. 4 (1982): 777-793.}
    \item{Loungani, Prakash, Mark Rush, and William Tave. "Stock market dispersion and unemployment." Journal of Monetary Economics 25, no. 3 (1990): 367-388.}
    \item{Maio, Paulo F. "Return dispersion and the predictability of stock returns." Available at SSRN 1986791 (2013).}
    \item{Maio, Paulo. "Cross-sectional return dispersion and the equity premium." Journal of Financial Markets 29 (2016): 87-109.}
    \item{Manela, Asaf, and Alan Moreira. "News implied volatility and disaster concerns." Journal of Financial Economics 123, no. 1 (2017): 137-162.}
    \item{Merton, Robert C. "An intertemporal capital asset pricing model." Econometrica 41, no. 5 (1973): 867-887.}
    \item{Novy-Marx, Robert. "Predicting anomaly performance with politics, the weather, global warming, sunspots, and the stars." Journal of Financial Economics 112, no. 2 (2014): 137-146.}
    \item{Pastor, Lubos, and Robert F. Stambaugh. "Liquidity risk and expected stock returns." Journal of Political economy 111, no. 3 (2003): 642-685.}
    \item{Stivers, Chris, and Licheng Sun. "Cross-sectional return dispersion and time variation in value and momentum premiums." Journal of Financial and Quantitative Analysis 45, no. 4 (2010): 987-1014.}
    \item{Verousis, Thanos, and Nikolaos Voukelatos. "Cross-sectional dispersion and expected returns." Quantitative finance 18, no. 5 (2018): 813-826.}

\end{enumerate}

\newpage

\newgeometry{left=2.5cm, right=0.75cm, top=1.75cm, bottom=1.25cm}
\section*{Appendix}

\begin{figure}[h!]
\textbf{Figure 1: Dispersion Measures}
\vskip 6 pt
\begin{flushleft}
{The plot describes the time series of cross-industry dispersion (CID) and within-industry dispersion (WID). All the measures are calculated as the mean absolute deviation of returns at the monthly frequency. Industries are defined according to Fama-French 49 industry classification. The paper uses value-weighted industry returns to calculate CID.}
\end{flushleft}
\centering
\vspace{0.64cm}
\includegraphics[width=0.9\textwidth]{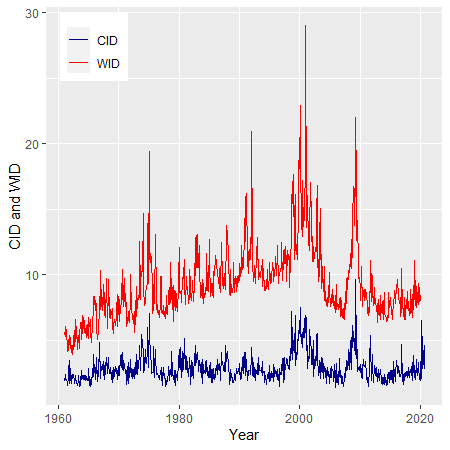}
\end{figure}

\begin{figure*}
\textbf{Figure 2: Performance of \$1 (log scale)}
\vskip 12 pt
\begin{flushleft}
{The plot describes the growth of \$1, invested in the long-short unisorted quintile value-weighted portfolio, formed on $\beta_{CID}$.}
\end{flushleft}
\centering
\includegraphics[width=1\textwidth]{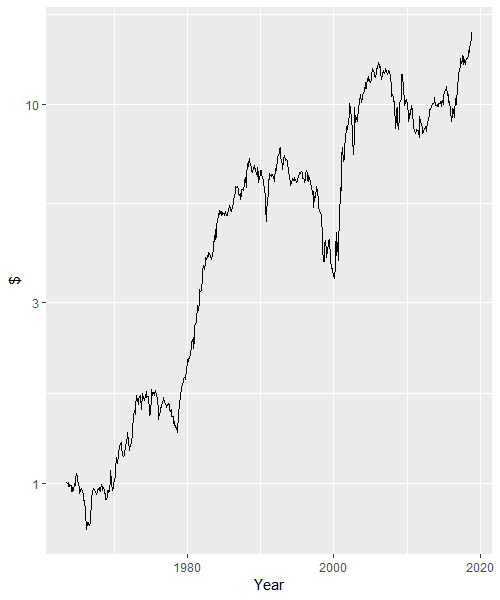}
\end{figure*}

\begin{figure*}
\textbf{Figure 3: Returns of quintile L/S portfolio using different Fama-French industry partitions}
\vskip 12 pt
\begin{flushleft}
{The plot describes monthly returns and abnormal returns of quintile L/S portfolios, formed on $\beta_{CID}$. X axis corresponds to the number of industries in industry classification. I use Fama-French industry definitions with different coarseness: 49, 30, 17, 10 and 5 industries to compute CID. To obtain finer sorts, I use SIC 2, 3 and 4 digits to classify the firms into industries. I compute CID only using the industries with at least 5 firms for every period. The median number of industries with at least 5 firms for SIC 2-4 classifications are 63, 174 and 212. In the case of SIC classifications, these numbers are depicted at x axis.  Abnormal returns are calculated with respect to Fama-French 5 factor model, augmented with Momentum and Short-term Reversal factors.}
\end{flushleft}
\centering
\includegraphics[width=1\textwidth]{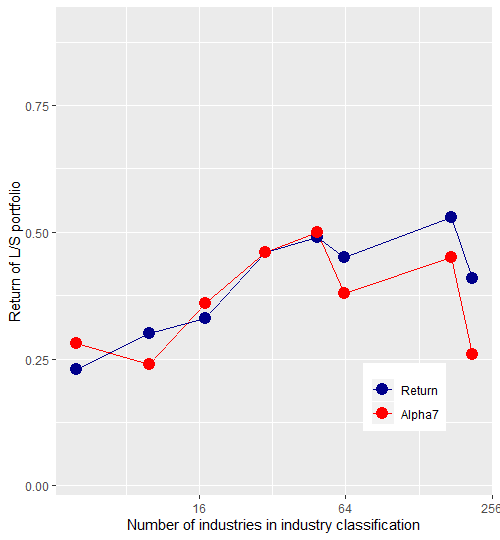}
\end{figure*}

\begin{figure*}
\textbf{Figure 4: Post-ranking betas vs pre-ranking betas using monthly returns}
\vskip 12 pt
\begin{flushleft}
{The plot describes realized full-sample $\beta_{CID}$ of quintile portfolios compared to ex-ante rolling betas $\hat{\beta}_{CID}$ of these portfolios.}
\end{flushleft}
\centering
\includegraphics[width=1\textwidth]{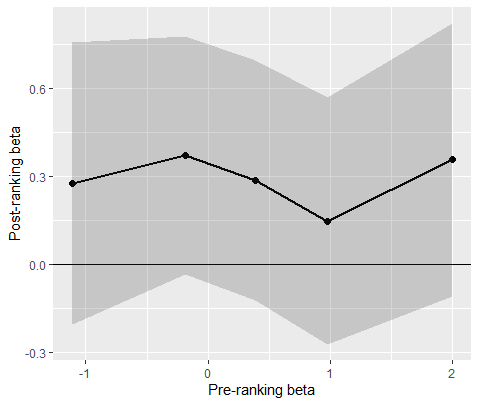}
\end{figure*}

\begin{figure*}
\textbf{Figure 5: Post-ranking betas vs pre-ranking betas using quarterly returns}
\vskip 12 pt
\begin{flushleft}
{The plot describes realized full-sample $\beta_{CID}$ of quintile portfolios compared to ex-ante rolling betas $\hat{\beta}_{CID}$ of these portfolios.}
\end{flushleft}
\centering
\includegraphics[width=1\textwidth]{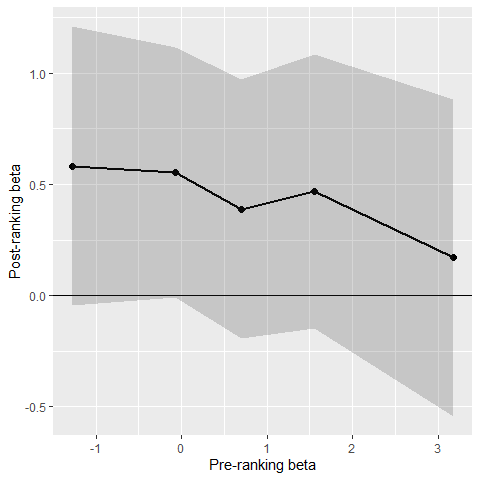}
\end{figure*}

\begin{figure*}
\textbf{Figure 6: Postranking daily $\beta_{CID}$}
\vskip 12 pt
\begin{flushleft}
{The plot describes the relationship between postranking and preranking $\beta_{CID}$ of decile portfolios, sorted on $\beta_{CID}$. Preranking betas are estimated from regression (3) using 504 most recent days. To calculate postranking betas, I run the  regression (3) for each decile portfolio.}
\end{flushleft}
\centering
\includegraphics[width=1\textwidth]{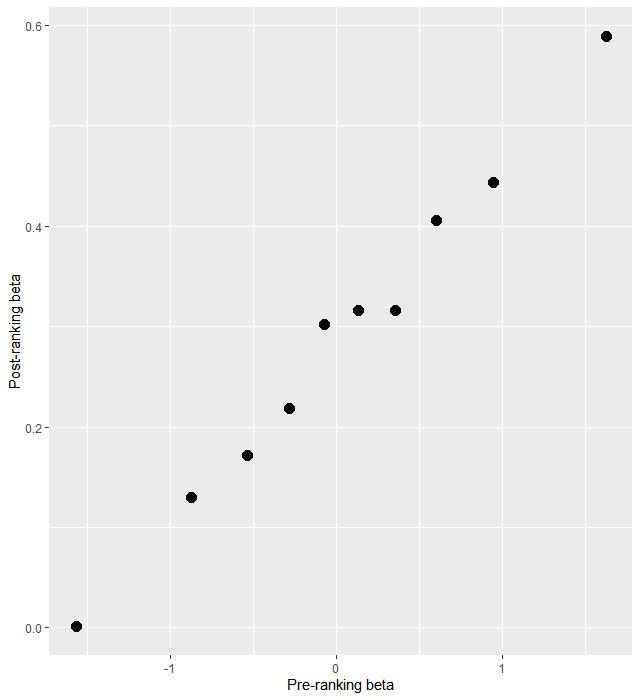}
\end{figure*}

\begin{figure*}
\textbf{Figure 7: Predictability of industry employment using industry returns}
\vskip 12 pt
\begin{flushleft}
{The plot shows T-statistics of regressor coefficient estimates from predictive panel regression for industry employment growth using lagged industry returns. Industry employment growth and return are normalized by aggregate employment growth and market return. Sample covers 1990-2019.}
\end{flushleft}
\centering
\includegraphics[width=1\textwidth]{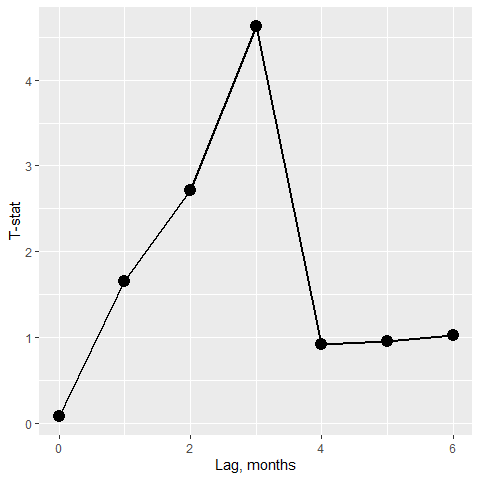}
\end{figure*}

\clearpage

\begin{table}[!htbp] \centering 
  \caption{\textbf{Correlations of differences in CID with differences in other variables}} 
  \label{} 
    \begin{flushleft}
    {\medskip\small
 The table reports correlations between differences in CID and changes in other variables at the monthly frequency. VIX is implied volatility, available starting from 1990. FU and MU are financial and macroeconomic uncertainty from Sydney Ludvigson website. VOL is the volatility of monthly value-weighted market index over the recent 24 months. CIV is common idosyncratic volatility (Kelly et al., 2016). }
    \medskip
    \end{flushleft}
\begin{tabular}{@{\extracolsep{5pt}} ccccccc} 
\\[-1.8ex]\hline 
\hline \\[-1.8ex] 
 & CID & FU & MU & VOL & CIV & VIX \\ 
\hline \\[-1.8ex] 
CID & $1$ & $0.24$ & $0.07$ & $0.30$ & $0.29$ & $0.11$ \\ 
FU & $0.24$ & $1$ & $0.44$ & $0.30$ & $0.39$ & $0.40$ \\ 
MU & $0.07$ & $0.44$ & $1$ & $0.11$ & $0.24$ & $0.36$ \\ 
VOL & $0.30$ & $0.30$ & $0.11$ & $1$ & $0.27$ & $0.34$ \\ 
CIV & $0.29$ & $0.39$ & $0.24$ & $0.27$ & $1$ & $0.50$ \\ 
VIX & $0.11$ & $0.40$ & $0.36$ & $0.34$ & $0.50$ & $1$ \\ 
\hline \\[-1.8ex] 
\end{tabular} 
\end{table}

\vspace{2cm}

\begin{table}[!htbp] \centering 
  \caption{\textbf{Characteristics of quintile $\beta_{CID}$-sorted vw portfolios}} 
  \label{} 
    \begin{flushleft}
    {\medskip\small
 The table reports mean values of characteristics of quintile portfolios, sorted on $\beta_{CID}$. Size is log of market equity in the previous month. B/M is Book-to-Market. OP is operating profitability. BA spread is the average bid-ask spread as a percentage of average price over the previous month. }
    \medskip
    \end{flushleft}
\begin{tabular}{@{\extracolsep{5pt}} lcccccc} 
\\[-1.8ex]\hline 
\hline \\[-1.8ex] 
 & Q1 & Q2 & Q3 & Q4 & Q5 & L/S \\ 
\hline \\[-1.8ex] 
Return & 0.79 & 0.63 & 0.58 & 0.45 & 0.30 & -0.49 \\ 
T-stat (Return) & $3.83$ & $3.51$ & $3.55$ & $2.59$ & $1.40$ & $-3.19$ \\ 
Prebeta & $-3.25$ & $-1.08$ & $0.27$ & $1.66$ & $4.12$ & $7.37$ \\ 
Size & $8.12$ & $8.76$ & $8.96$ & $8.87$ & $8.40$ & $0.28$ \\ 
log(B/M) & $-0.78$ & $-0.77$ & $-0.78$ & $-0.80$ & $-0.84$ & $-0.07$ \\ 
OP & $0.16$ & $0.17$ & $0.17$ & $0.17$ & $0.17$ & $0.01$ \\ 
Investment & $0.15$ & $0.15$ & $0.14$ & $0.15$ & $0.22$ & $0.08$ \\ 
Beta & $1.04$ & $0.98$ & $0.99$ & $1.03$ & $1.14$ & $0.11$ \\ 
BA Spread & $0.31$ & $0.23$ & $0.20$ & $0.19$ & $0.21$ & $-0.10$ \\ 
Momentum 12-2 & $0.12$ & $0.09$ & $0.10$ & $0.12$ & $0.19$ & $0.07$ \\ 
Volatility (1m) & $2.01$ & $1.68$ & $1.61$ & $1.69$ & $2.07$ & $0.06$ \\ 
Volatility (12m) & $2.13$ & $1.75$ & $1.69$ & $1.79$ & $2.21$ & $0.08$ \\ 
\hline \\[-1.8ex] 
\end{tabular} 
\end{table}

\begin{table}[!htbp] \centering 
  \caption{\textbf{Returns of quintile $\beta_{CID}$-sorted portfolios}} 
  \label{} 
  \begin{flushleft}
    {\medskip\small
 The table reports mean monthly excess returns of quintile portfolios, sorted on $\beta_{CID}$. Q1 is the quintile portfolio with the lowest $\beta_{CID}$ and Q5 contains the highest $\beta_{CID}$ stocks. L/S is the long-short portfolio, formed by buying Q5 and selling Q1. Value-weighted portfolios are constructed using as weights the market capitalization from the previous month. The returns are calculated at the monthly frequency over 1963-2018.}
    \medskip
    \end{flushleft}
\begin{tabular}{@{\extracolsep{5pt}} ccccccc} 
\\[-1.8ex]\hline 
\hline \\[-1.8ex] 
 & Q1 & Q2 & Q3 & Q4 & Q5 & L/S \\ 
\hline \\[-1.8ex] 
Mean ew & 0.81 & 0.79$^{***}$ & 0.73$^{***}$ & 0.65$^{***}$ & 0.50$^{**}$ & -0.31$^{**}$ \\ 
T-stat ew & 3.56 & 4.09 & 3.96 & 3.42 & 2.09 & -2.47 \\ 
Mean vw & 0.79 & 0.63$^{***}$ & 0.58$^{***}$ & 0.45$^{***}$ & 0.30 & -0.49$^{***}$ \\ 
T-stat vw & 3.83 & 3.51 & 3.55 & 2.59 & 1.40 & -3.19 \\ 
\hline \\[-1.8ex] 
\end{tabular} 
\end{table}

\begin{table}[!htbp] \centering 
  \caption{\textbf{Abnormal returns of quintile $\beta_{CID}$-sorted vw portfolios}} 
  \label{} 
  \begin{flushleft}
    {\medskip\small
 The table reports abnormal monthly returns of the long-short value-weighted quintile portfolio, formed from sorts on $\beta_{CID}$. The last column contains the abnormal returns with respect to Fama-French 5 factor model, augmented with momentum and short-term reversal factors. The returns are calculated at the monthly frequency over 1963-2018.}
    \medskip
    \end{flushleft}
\begin{tabular}{@{\extracolsep{5pt}} lcccccc} 
\\[-1.8ex]\hline 
\hline \\[-1.8ex] 
Statistic & Ret & $\alpha_{CAPM}$ & $\alpha_{FF3}$ & $\alpha_{Carhart}$ & $\alpha_{FF5}$ & $\alpha_{FF5+UMD+STR}$ \\ 
\hline \\[-1.8ex] 
LS & -0.49$^{***}$ & -0.52$^{***}$ & -0.40$^{***}$ & -0.63$^{***}$ & -0.29$^{*}$ & -0.50$^{***}$ \\ 
 & [ -3.19] & [ -3.42] & [ -2.64] & [ -4.22] & [ -1.87] & [ -3.26] \\ 
\hline \\[-1.8ex] 
\end{tabular} 
\end{table}

\begin{table}[!htbp] \centering 
  \caption{\textbf{Factor loadings of quintile $\beta_{CID}$-sorted vw portfolios}} 
  \label{} 
  \begin{flushleft}
    {\medskip\small
 The table reports factor loadings of value-weighted portfolios, sorted on $\beta_{CID}$, on Fama-French 5 factor model, augmented with momentum and short-term reversal factors. The returns are calculated at the monthly frequency over 1963-2018.}
    \medskip
    \end{flushleft}
\begin{tabular}{@{\extracolsep{0pt}} ccccccccccc} 
\\[-1.8ex]\hline 
\hline \\[-1.8ex] 
Quintile & Ret & Alpha & EMKT & HML & SMB & RMW & CMA & MOM & STR & adjR2 \\ 
\hline \\[-1.8ex] 
1 & 0.79 & 0.27 & 1.07 & 0.10 & 0.17 & 0.06 & -0.01 & -0.16 & -0.01 & 0.85 \\ 
 & [ 3.83] & [ 3.07] & [ 49.36] & [ 2.34] & [ 5.78] & [ 1.50] & [ -0.16] & [ -7.44] & [ -0.29] &  \\ 
2 & 0.63 & 0.07 & 1.02 & 0.13 & -0.01 & 0.19 & 0.08 & -0.12 & 0.02 & 0.90 \\ 
 & [ 3.51] & [ 1.08] & [ 67.00] & [ 4.46] & [ -0.49] & [ 6.32] & [ 1.89] & [ -8.30] & [ 0.86] &  \\ 
3 & 0.58 & 0.03 & 0.97 & 0.05 & -0.07 & 0.18 & 0.11 & -0.05 & 0.01 & 0.94 \\ 
 & [ 3.55] & [ 0.64] & [ 91.19] & [ 2.27] & [ -4.64] & [ 8.87] & [ 3.79] & [ -5.02] & [ 0.87] &  \\ 
4 & 0.45 & -0.09 & 0.99 & -0.04 & -0.05 & 0.06 & 0.04 & 0.02 & 0.03 & 0.92 \\ 
 & [ 2.59] & [ -1.72] & [ 78.18] & [ -1.47] & [ -2.82] & [ 2.32] & [ 1.00] & [ 1.47] & [ 1.77] &  \\ 
5 & 0.30 & -0.23 & 1.11 & -0.04 & 0.08 & -0.28 & -0.18 & 0.13 & 0.00 & 0.87 \\ 
 & [ 1.40] & [ -2.72] & [ 53.56] & [ -0.95] & [ 2.70] & [ -7.08] & [ -3.09] & [ 6.54] & [ -0.11] &  \\ 
LS & -0.49 & -0.50 & 0.04 & -0.14 & -0.10 & -0.35 & -0.17 & 0.29 & 0.01 & 0.14 \\ 
 & [ -3.19] & [ -3.26] & [ 0.93] & [ -1.87] & [ -1.85] & [ -4.75] & [ -1.60] & [ 7.88] & [ 0.11] &  \\ 
\hline \\[-1.8ex] 
\end{tabular} 
\end{table} 

\vspace{1cm}

\begin{table}[!htbp] \centering 
  \caption{\textbf{Abnormal returns of 2x3 double-sorted portfolios on size and $\beta_{CID}$}}
  \label{} 
  \begin{flushleft}
    {\medskip\small
 The table reports abnormal monthly returns of long-short value-weighted portfolios, formed from independent 2 by 3 double sorts on size and $\beta_{CID}$. By convention, in tercile sort I use 30 and 70 percentiles. The long-short portfolios are formed in the following way:
 $L/S Size = \frac{1}{3}(SmallLow+SmallMedium+SmallHigh) - \frac{1}{3}(BigLow+BigMedium+BigHigh)$,
 $L/S CID = \frac{1}{2}(SmallLow+BigLow) - \frac{1}{2}(SmallHigh+BigHigh)$. \\
 The last column contains the abnormal returns with respect to Fama-French 5 factor model, augmented with momentum and short-term reversal factors. The returns are calculated at the monthly frequency over 1963-2018.}
    \medskip
    \end{flushleft}
\begin{tabular}{@{\extracolsep{5pt}} lcccccc} 
\\[-1.8ex]\hline 
\hline \\[-1.8ex] 
Statistic & Ret & $\alpha_{CAPM}$ & $\alpha_{FF3}$ & $\alpha_{Carhart}$ & $\alpha_{FF5}$ & $\alpha_{FF5+UMD+STR}$ \\ 
\hline \\[-1.8ex] 
L/S Size & 0.21$^{**}$ & 0.14 & -0.04 & 0.01 & -0.07$^{**}$ & -0.05$^{*}$ \\ 
T-stat & [ 2.11] & [ 1.45] & [ -1.15] & [ 0.36] & [ -2.24] & [ -1.66] \\ 
L/S CID & 0.28$^{***}$ & 0.30$^{***}$ & 0.20$^{**}$ & 0.40$^{***}$ & 0.17$^{*}$ & 0.33$^{***}$ \\ 
T-stat & [ 2.72] & [ 2.94] & [ 2.05] & [ 4.16] & [ 1.68] & [ 3.40] \\ 
\hline \\[-1.8ex] 
\end{tabular}
\end{table}

\begin{table}[!htbp] \centering 
  \caption{\textbf{Fama-MacBeth regression}} 
  \label{} 
  \begin{flushleft}
    {\medskip\small
The table reports the results of Fama-MacBeth regression of excess returns on characteristics of decile $\beta_{CID}$-sorted portfolios. Size is log(ME) in the previous month. Momentum is the return over the past year, excluding the most recent month. Investment is defined as the growth in total assets over the recent year. $\beta_{VOL}$ is the sensitivity to the volatility of monthly market returns over the recent 2 years. The returns are calculated at the monthly frequency over 1963-2018.}
    \medskip
    \end{flushleft}
\begin{tabular}{@{\extracolsep{5pt}}lccccc} 
\\[-1.8ex]\hline 
\hline \\[-1.8ex] 
 & \multicolumn{5}{c}{\textit{Dependent variable: Return}} \\ 
\cline{2-6} 
\\[-1.8ex] & (1) & (2) & (3) & (4) & (5)\\ 
\hline \\[-1.8ex] 
 $\beta_{CID}$ & $-$0.12$^{***}$ & $-$0.09$^{***}$ & $-$0.09$^{***}$ & $-$0.08$^{**}$ & $-$0.10$^{**}$ \\ 
  & [$-$3.99] & [$-$3.43] & [$-$3.46] & [$-$2.54] & [$-$2.05] \\ 
  & & & & & \\ 
 $\beta$ &  & 0.44$^{**}$ & 0.28 & $-$0.38 & $-$0.60 \\ 
  &  & [2.18] & [0.62] & [$-$0.74] & [$-$0.84] \\ 
  & & & & & \\ 
 size &  &  & 0.01 & 0.06 & 0.01 \\ 
  &  &  & [0.30] & [1.19] & [0.14] \\ 
  & & & & & \\ 
 logbm &  &  & 0.03 & 0.12 & 0.18 \\ 
  &  &  & [0.16] & [0.62] & [0.61] \\ 
  & & & & & \\ 
 mom122 &  &  &  & 1.24$^{*}$ & 1.50 \\ 
  &  &  &  & [1.69] & [1.52] \\ 
  & & & & & \\ 
 op &  &  &  &  & 3.19 \\ 
  &  &  &  &  & [1.38] \\ 
  & & & & & \\ 
 inv &  &  &  &  & $-$1.12 \\ 
  &  &  &  &  & [$-$0.74] \\ 
  & & & & & \\ 
 Constant & 0.52$^{***}$ & 0.09 & 0.10$^{*}$ & 0.09$^{*}$ & 0.08$^{**}$ \\ 
  & [3.22] & [1.19] & [1.80] & [1.96] & [2.47] \\ 
  & & & & & \\ 
\hline \\[-1.8ex] 
Observations & 10 & 10 & 10 & 10 & 10 \\ 
Adjusted R$^{2}$ & 0.38 & 0.58 & 0.75 & 0.80 & 0.89 \\ 
\hline 
\hline \\[-1.8ex] 
\textit{Note:}  & \multicolumn{5}{r}{$^{*}$p$<$0.1; $^{**}$p$<$0.05; $^{***}$p$<$0.01} \\ 
\end{tabular}
\end{table}

\clearpage

\begin{table}[!htbp] \centering 
  \caption{\textbf{Abnormal returns of 5x5 portfolios, double-sorted on within-industry dispersion $\beta_{WID}$ and $\beta_{CID}$}} 
  \label{} 
  \begin{flushleft}
    {\medskip\small
 The table reports abnormal monthly returns of long-short value-weighted portfolios, formed from independent 5 by 5 double sorts on $\beta_{WID}$ and $\beta_{CID}$. WID (within-industry dispersion) is a mean absolute deviation of returns of the stocks within each industry, averaged across 49 industries. The long-short portfolios are formed in the following way: \\
 \scriptsize
 \vspace{0.1cm}
 $L/S WID = \frac{1}{5}(LowWIDLowCID+LowWIDCID2+LowWIDCID3+LowWIDCID4+LowWIDHighCID) - \frac{1}{5}(HighWIDLowCID+HighWIDCID2+HighWIDCID3+HighWIDCID4+HighWIDHighCID)$, \\
 $L/S CID = \frac{1}{5}(LowCIDLowWID+LowCIDWID2+LowCIDWID3+LowCIDWID4+LowCIDHighWID) - \frac{1}{5}(HighCIDLowWID+HighCIDWID2+HighCIDWID3+HighCIDWID4+HighCIDHighWID)$. \\
 \normalsize
 The last column contains the abnormal returns with respect to Fama-French 5 factor model, augmented with momentum and short-term reversal factors. The returns are calculated at the monthly frequency over 1963-2018.}
    \medskip
    \end{flushleft}
\begin{tabular}{@{\extracolsep{5pt}} lcccccc} 
\\[-1.8ex]\hline 
\hline \\[-1.8ex] 
Statistic & Ret & $\alpha_{CAPM}$ & $\alpha_{FF3}$ & $\alpha_{Carhart}$ & $\alpha_{FF5}$ & $\alpha_{FF5+UMD+STR}$ \\ 
\hline \\[-1.8ex] 
L/S WID & 0.07 & 0.32$^{*}$ & 0.20 & 0.21 & -0.15 & -0.10 \\ 
T-stat & [ 0.40] & [ 1.74] & [ 1.22] & [ 1.28] & [ -0.97] & [ -0.63] \\ 
L/S CID & 0.30$^{**}$ & 0.23 & 0.14 & 0.27$^{*}$ & 0.20 & 0.28$^{**}$ \\ 
T-stat & [ 2.22] & [ 1.58] & [ 1.02] & [ 1.91] & [ 1.42] & [ 1.97] \\ 
\hline \\[-1.8ex] 
\end{tabular} 
\end{table}

\vspace{2cm}

\begin{table}[!htbp] \centering 
  \caption{\textbf{Abnormal returns of 5x5 portfolios, double-sorted on cross-sectional dispersion $\beta_{CSD}$ and $\beta_{CID}$}} 
  \label{} 
  \begin{flushleft}
    {\medskip\small
 The table reports abnormal monthly returns of long-short value-weighted portfolios, formed from independent 5 by 5 double sorts on $\beta_{CSD}$ and $\beta_{CID}$. CSD (cross-sectional dispersion) is a mean absolute deviation of returns of all stocks. The long-short portfolios are formed in the following way: \\
  \scriptsize
  \vspace{0.1cm}
$L/S CSD = \frac{1}{5}(LowCSDLowCID+LowCSDCID2+LowCSDCID3+LowCSDCID4+LowCSDHighCID) - \frac{1}{5}(HighCSDLowCID+HighCSDCID2+HighCSDCID3+HighCSDCID4+HighCSDHighCID)$, \\
$L/S CID = \frac{1}{5}(LowCIDLowCSD+LowCIDCSD2+LowCIDCSD3+LowCIDCSD4+LowCIDHighCSD) - \frac{1}{5}(HighCIDLowCSD+HighCIDCSD2+HighCIDCSD3+HighCIDCSD4+HighCIDHighCSD)$. \\
\normalsize
 The last column contains the abnormal returns with respect to Fama-French 5 factor model, augmented with momentum and short-term reversal factors. The returns are calculated at the monthly frequency over 1963-2018.}
    \medskip
    \end{flushleft}
\begin{tabular}{@{\extracolsep{5pt}} lcccccc} 
\\[-1.8ex]\hline 
\hline \\[-1.8ex] 
Statistic & Ret & $\alpha_{CAPM}$ & $\alpha_{FF3}$ & $\alpha_{Carhart}$ & $\alpha_{FF5}$ & $\alpha_{FF5+UMD+STR}$ \\ 
\hline \\[-1.8ex] 
L/S CSD & 0.10 & 0.35$^{*}$ & 0.21 & 0.23 & -0.21 & -0.17 \\ 
T-stat & [ 0.50] & [ 1.80] & [ 1.23] & [ 1.30] & [ -1.28] & [ -1.03] \\ 
L/S CID & 0.29$^{**}$ & 0.21 & 0.12 & 0.26$^{*}$ & 0.22 & 0.30$^{**}$ \\ 
T-stat & [ 2.11] & [ 1.46] & [ 0.87] & [ 1.82] & [ 1.50] & [ 2.09] \\ 
\hline \\[-1.8ex] 
\end{tabular} 
\end{table}

\begin{table}[!htbp] \centering 
  \caption{\textbf{Abnormal returns of 5x5 portfolios, double-sorted on $\beta_{CID}$ and other variables}}
  \label{} 
  \begin{flushleft}
    {\medskip\small
 Construction procedure is equivalent to the one used in the previous tables. MU and FU are macroeconomic and financial uncertainty indices from Ludvigson et al. (2015). Market volatility is a standard deviation of monthly value-weighted market returns over the last 24 months. }
    \medskip
    \end{flushleft}
    
\begin{tabularx}{\linewidth}{p{2cm}p{1.5cm}p{1.5cm}p{1.5cm}p{1.5cm}p{1.5cm}p{1.5cm}}
    \toprule
    \multicolumn{7}{l}{\textbf{Panel A: Market volatility vs CID}} \\
    \midrule 
\\[-1.8ex]\hline 
\hline \\[-1.8ex] 
Statistic & Ret & $\alpha_{CAPM}$ & $\alpha_{FF3}$ & $\alpha_{Carhart}$ & $\alpha_{FF5}$ & $\alpha_{FF5+UMD+STR}$ \\ 
\hline \\[-1.8ex] 
L/S VOL & 0.26$^{*}$ & 0.14 & 0.19 & 0.22 & 0.38$^{***}$ & 0.36$^{***}$ \\ 
T-stat & [ 1.87] & [ 1.06] & [ 1.42] & [ 1.61] & [ 2.80] & [ 2.61] \\ 
L/S CID & 0.42$^{***}$ & 0.47$^{***}$ & 0.38$^{***}$ & 0.60$^{***}$ & 0.34$^{**}$ & 0.52$^{***}$ \\ 
T-stat & [ 3.09] & [ 3.43] & [ 2.82] & [ 4.55] & [ 2.41] & [ 3.81] \\ 
\hline \\[-1.8ex]
\end{tabularx}

\begin{tabularx}{\linewidth}{p{2cm}p{1.5cm}p{1.5cm}p{1.5cm}p{1.5cm}p{1.5cm}p{1.5cm}}
    \toprule
    \multicolumn{7}{l}{\textbf{Panel B: Macroeconomic uncertainty (Ludvigson 2015) vs CID}} \\
    \midrule  
\\[-1.8ex]\hline 
\hline \\[-1.8ex] 
Statistic & Ret & $\alpha_{CAPM}$ & $\alpha_{FF3}$ & $\alpha_{Carhart}$ & $\alpha_{FF5}$ & $\alpha_{FF5+UMD+STR}$ \\ 
\hline \\[-1.8ex] 
L/S MU & -0.16 & -0.26$^{*}$ & -0.22 & -0.17 & 0.00 & -0.07 \\ 
T-stat & [ -1.09] & [ -1.83] & [ -1.58] & [ -1.25] & [ 0.02] & [ -0.51] \\ 
L/S CID & 0.41$^{***}$ & 0.43$^{***}$ & 0.28$^{**}$ & 0.54$^{***}$ & 0.18 & 0.41$^{***}$ \\ 
T-stat & [ 2.83] & [ 2.97] & [ 1.96] & [ 3.98] & [ 1.24] & [ 3.00] \\ 
\hline \\[-1.8ex] 
\end{tabularx} 

\begin{tabularx}{\linewidth}{p{2cm}p{1.5cm}p{1.5cm}p{1.5cm}p{1.5cm}p{1.5cm}p{1.5cm}}
    \toprule
    \multicolumn{7}{l}{\textbf{Panel C: Financial uncertainty (Ludvigson 2015) vs CID}} \\
    \midrule 
\\[-1.8ex]\hline 
\hline \\[-1.8ex] 
Statistic & Ret & $\alpha_{CAPM}$ & $\alpha_{FF3}$ & $\alpha_{Carhart}$ & $\alpha_{FF5}$ & $\alpha_{FF5+UMD+STR}$ \\ 
\hline \\[-1.8ex] 
L/S FU & -0.04 & -0.16 & -0.29$^{**}$ & -0.40$^{***}$ & -0.41$^{***}$ & -0.48$^{***}$ \\ 
T-stat & [ -0.30] & [ -1.10] & [ -2.07] & [ -2.89] & [ -2.91] & [ -3.33] \\ 
L/S CID & 0.44$^{***}$ & 0.49$^{***}$ & 0.39$^{***}$ & 0.67$^{***}$ & 0.41$^{***}$ & 0.65$^{***}$ \\ 
T-stat & [ 3.33] & [ 3.66] & [ 2.97] & [ 5.34] & [ 3.02] & [ 5.08] \\ 
\hline \\[-1.8ex] 
\end{tabularx}

\begin{tabularx}{\linewidth}{p{2cm}p{1.5cm}p{1.5cm}p{1.5cm}p{1.5cm}p{1.5cm}p{1.5cm}}
    \toprule
    \multicolumn{7}{l}{\textbf{Panel D: VIX vs CID}} \\
    \midrule  
\\[-1.8ex]\hline 
\hline \\[-1.8ex] 
Statistic & Ret & $\alpha_{CAPM}$ & $\alpha_{FF3}$ & $\alpha_{Carhart}$ & $\alpha_{FF5}$ & $\alpha_{FF5+UMD+STR}$ \\ 
\hline \\[-1.8ex] 
L/S VIX & -0.09 & -0.50$^{**}$ & -0.47$^{**}$ & -0.34$^{*}$ & -0.15 & -0.07 \\ 
T-stat & [ -0.36] & [ -2.54] & [ -2.46] & [ -1.80] & [ -0.80] & [ -0.38] \\ 
L/S CID & 0.13 & 0.17 & 0.05 & 0.28 & -0.11 & 0.07 \\ 
T-stat & [ 0.64] & [ 0.83] & [ 0.24] & [ 1.55] & [ -0.57] & [ 0.40] \\ 
\hline \\[-1.8ex] 
\end{tabularx} 

\end{table}

\begin{table}[!htbp] \centering 
  \caption{\textbf{Abnormal returns of 5x5 portfolios, double-sorted on $\beta_{CID}$ and $\beta_{CIV}$}}
  \label{} 
  \begin{flushleft}
    {\medskip\small
 The table reports abnormal monthly returns of long-short value-weighted portfolios, formed from independent 5 by 5 double sorts on $\beta_{CID}$ and $\beta_{CIV}$. CIV is common idiosyncratic volatility from Herskovic, Kelly, Lustig and Van Nieuwerburgh (2016).  The long-short portfolios are formed in the following way: \\
   \scriptsize
  \vspace{0.32cm}
$L/S CIV = \frac{1}{5}(LowCIVLowCID+LowCIVCID2+LowCIVCID3+LowCIVCID4+LowCIVHighCID) - \frac{1}{5}(HighCIVLowCID+HighCIVCID2+HighCIVCID3+HighCIVCID4+HighCIVHighCID)$, \\
$L/S CID = \frac{1}{5}(LowCIDLowCIV+LowCIDCIV2+LowCIDCIV3+LowCIDCIV4+LowCIDHighCIV) - \frac{1}{5}(HighCIDLowCIV+HighCIDCIV2+HighCIDCIV3+HighCIDCIV4+HighCIDHighCIV)$. \\
 }
    \medskip
    \end{flushleft}
\begin{tabular}{@{\extracolsep{5pt}} lcccccc} 
\\[-1.8ex]\hline 
\hline \\[-1.8ex] 
Statistic & Ret & $\alpha_{CAPM}$ & $\alpha_{FF3}$ & $\alpha_{Carhart}$ & $\alpha_{FF5}$ & $\alpha_{FF5+UMD+STR}$ \\ 
\hline \\[-1.8ex] 
L/S CIV & 0.08 & -0.01 & -0.01 & -0.05 & 0.05 & 0.01 \\ 
T-stat & [ 0.57] & [ -0.07] & [ -0.10] & [ -0.37] & [ 0.39] & [ 0.04] \\ 
L/S CID & 0.40$^{***}$ & 0.43$^{***}$ & 0.28$^{**}$ & 0.54$^{***}$ & 0.20 & 0.43$^{***}$ \\ 
T-stat & [ 2.78] & [ 2.94] & [ 2.02] & [ 4.02] & [ 1.40] & [ 3.16] \\ 
\hline \\[-1.8ex] 
\end{tabular} 
\end{table}

\begin{table}[!htbp] \centering 
  \caption{\textbf{Spanning tests between CID factor and CIV factor}} 
  \label{} 
  \begin{flushleft}
    {\medskip\small
The table reports the results of spanning tests between CID factor and CIV factor. Both factors are constructed as the long-short quintile portfolios from the sorts on $\beta_{CID}$ and $\beta_{CIV}$ respectively. The both factors are long stocks with high sensitivity to non-traded factor and short stocks with low sensitivity. $\beta_{CIV}$ is the sensitivity to the common idiosyncratic volatility CIV (Herskovic et al., 2016), estimated over the recent 5 years. The returns are calculated at the monthly frequency over 1963-2018.}
    \medskip
    \end{flushleft}
\begin{tabular}{@{\extracolsep{5pt}}lcccc} 
\\[-1.8ex]\hline 
\hline \\[-1.8ex] 
 & \multicolumn{4}{c}{\textit{Dependent variable:}} \\ 
\cline{2-5} 
\\[-1.8ex] & \multicolumn{2}{c}{CID factor} & \multicolumn{2}{c}{CIV factor} \\ 
\\[-1.8ex] & (1) & (2) & (3) & (4)\\ 
\hline \\[-1.8ex] 
 EMKT & 0.033 & 0.066$^{*}$ & $-$0.154$^{***}$ & $-$0.160$^{***}$ \\ 
  & [0.926] & [1.865] & [$-$4.354] & [$-$4.651] \\ 
  & & & & \\ 
 SMB & $-$0.076 & $-$0.027 & $-$0.231$^{***}$ & $-$0.215$^{***}$ \\ 
  & [$-$1.479] & [$-$0.519] & [$-$4.520] & [$-$4.292] \\ 
  & & & & \\ 
 HML & $-$0.191$^{***}$ & $-$0.164$^{**}$ & $-$0.129$^{*}$ & $-$0.088 \\ 
  & [$-$2.667] & [$-$2.328] & [$-$1.805] & [$-$1.258] \\ 
  & & & & \\ 
 RMW & $-$0.227$^{***}$ & $-$0.205$^{***}$ & $-$0.102 & $-$0.054 \\ 
  & [$-$3.206] & [$-$2.957] & [$-$1.453] & [$-$0.780] \\ 
  & & & & \\ 
 CMA & $-$0.131 & $-$0.157 & 0.117 & 0.145 \\ 
  & [$-$1.243] & [$-$1.513] & [1.119] & [1.413] \\ 
  & & & & \\ 
 MOM & 0.284$^{***}$ & 0.270$^{***}$ & 0.067$^{**}$ & 0.007 \\ 
  & [8.289] & [8.020] & [1.977] & [0.205] \\ 
  & & & & \\ 
 CIVf &  & 0.215$^{***}$ &  &  \\ 
  &  & [5.326] &  &  \\ 
  & & & & \\ 
 CIDf &  &  &  & 0.211$^{***}$ \\ 
  &  &  &  & [5.326] \\ 
  & & & & \\ 
 Constant & $-$0.501$^{***}$ & $-$0.468$^{***}$ & $-$0.152 & $-$0.046 \\ 
  & [$-$3.392] & [$-$3.239] & [$-$1.040] & [$-$0.321] \\ 
  & & & & \\ 
\hline \\[-1.8ex] 
Observations & 605 & 605 & 605 & 605 \\ 
R$^{2}$ & 0.174 & 0.212 & 0.102 & 0.143 \\ 
Adjusted R$^{2}$ & 0.166 & 0.202 & 0.093 & 0.133 \\ 
\hline 
\hline \\[-1.8ex]
\textit{Note:}  & \multicolumn{4}{r}{$^{*}$p$<$0.1; $^{**}$p$<$0.05; $^{***}$p$<$0.01} \\ 
\end{tabular} 
\end{table}

\begin{table}[!htbp] \centering 
  \caption{\textbf{Predictive regressions for unemployment}} 
  \label{} 
        \begin{flushleft}
    {\medskip\small
 The table reports the results of the regression, predicting seasonally-adjusted unemployment at the quarterly frequency. Mkt and Vol correspond to the value-weighted market return and monthly volatility of value-weighted market returns over most recent 24 months. All dependent variables are differenced using the specification from Pastor and Stambaugh (2003). Long-term (LT) unemployment is defined as unemployment for more than 15 weeks. Short-term (ST) unemployment is unemployment for less than 5 weeks. The sample is restricted to 1948-2019 due to availability of unemployment data. I use Newey-West standard errors with 4 lags.}
    \medskip
    \end{flushleft}
\begin{tabular}{@{\extracolsep{5pt}}lcccccc} 
\\[-1.8ex]\hline 
\hline \\[-1.8ex] 
 & \multicolumn{6}{c}{\textit{Dependent variable:}} \\ 
\cline{2-7} 
 & \multicolumn{2}{c}{Unemployment} & \multicolumn{2}{c}{LT Unemployment} & \multicolumn{2}{c}{ST Unemployment} \\ 
\\[-1.8ex] & (1) & (2) & (3) & (4) & (5) & (6)\\ 
\hline \\[-1.8ex] 
 CID & 4.01$^{***}$ & 3.95$^{***}$ & 2.65$^{**}$ & 2.88$^{***}$ & 1.22$^{*}$ & 0.90 \\ 
  & [2.77] & [2.88] & [2.15] & [2.62] & [1.79] & [1.59] \\ 
  & & & & & & \\ 
 Mkt &  & $-$0.82$^{***}$ &  & $-$0.03 &  & $-$0.80$^{***}$ \\ 
  &  & [$-$2.58] &  & [$-$0.15] &  & [$-$4.11] \\ 
  & & & & & & \\ 
 Vol &  & 0.18$^{**}$ &  & 0.10$^{**}$ &  & 0.07$^{*}$ \\ 
  &  & [2.40] &  & [2.18] &  & [1.88] \\ 
  & & & & & & \\ 
 Constant & 0.004 & $-$0.01 & 0.01 & $-$0.002 & $-$0.002 & $-$0.01 \\ 
  & [0.09] & [$-$0.27] & [0.19] & [$-$0.07] & [$-$0.08] & [$-$0.49] \\ 
  & & & & & & \\ 
\hline \\[-1.8ex] 
Observations & 286 & 220 & 286 & 220 & 286 & 220 \\ 
R$^{2}$ & 0.03 & 0.19 & 0.05 & 0.14 & 0.01 & 0.18 \\ 
Adjusted R$^{2}$ & 0.03 & 0.18 & 0.04 & 0.13 & 0.004 & 0.17 \\ 
\hline 
\hline \\[-1.8ex] 
\textit{Note:}  & \multicolumn{6}{r}{$^{*}$p$<$0.1; $^{**}$p$<$0.05; $^{***}$p$<$0.01} \\ 
\end{tabular} 
\end{table}

\begin{table}[!htbp] \centering 
  \caption{\textbf{Abnormal returns of 5x5 portfolios, double-sorted on within-industry dispersion $\beta_{WID}$ and $\beta_{CID}$} using FF5 industry classification} 
  \label{} 
  \begin{flushleft}
    {\medskip\small
 The table reports abnormal monthly returns of long-short value-weighted portfolios, formed from independent 5 by 5 double sorts on $\beta_{WID}$ and $\beta_{CID}$. WID (within-industry dispersion) is a mean absolute deviation of returns of the stocks within each industry, averaged across 5 industries. The long-short portfolios are formed in the following way: \\
 \scriptsize
 \vspace{0.1cm}
 $L/S WID = \frac{1}{5}(LowWIDLowCID+LowWIDCID2+LowWIDCID3+LowWIDCID4+LowWIDHighCID) - \frac{1}{5}(HighWIDLowCID+HighWIDCID2+HighWIDCID3+HighWIDCID4+HighWIDHighCID)$, \\
 $L/S CID = \frac{1}{5}(LowCIDLowWID+LowCIDWID2+LowCIDWID3+LowCIDWID4+LowCIDHighWID) - \frac{1}{5}(HighCIDLowWID+HighCIDWID2+HighCIDWID3+HighCIDWID4+HighCIDHighWID)$. \\
 \normalsize
 The last column contains the abnormal returns with respect to Fama-French 5 factor model, augmented with momentum and short-term reversal factors. The returns are calculated at the monthly frequency over 1963-2018.}
    \medskip
    \end{flushleft}
\begin{tabular}{@{\extracolsep{5pt}} ccccccc} 
\\[-1.8ex]\hline 
\hline \\[-1.8ex] 
Statistic & Ret & $\alpha_{CAPM}$ & $\alpha_{FF3}$ & $\alpha_{Carhart}$ & $\alpha_{FF5}$ & $\alpha_{FF5+UMD+STR}$ \\ 
\hline \\[-1.8ex] 
L/S WID & 0.10 & 0.30$^{**}$ & 0.23$^{*}$ & 0.16 & -0.02 & -0.05 \\ 
T-stat & [ 0.61] & [ 2.03] & [ 1.72] & [ 1.20] & [ -0.13] & [ -0.39] \\
L/S CID & 0.17 & 0.13 & -0.01 & 0.27$^{***}$ & 0.09 & 0.27$^{**}$ \\ 
T-stat & [ 1.44] & [ 1.14] & [ -0.08] & [ 2.63] & [ 0.77] & [ 2.54] \\ 
\hline \\[-1.8ex]
\end{tabular} 
\end{table}

\begin{table}[!htbp] \centering 
  \caption{\textbf{Returns of the quintile portfolios, formed on $\beta_{CID}$, calculated from abnormal returns of FF49 industry portfolios}} 
  \label{} 
    \begin{flushleft}
    {\medskip\small
 The table reports mean monthly excess returns of quintile portfolios, sorted on $\beta_{CID}$. Q1 is the quintile portfolio with the lowest $\beta_{CID}$ and Q5 contains the highest $\beta_{CID}$ stocks. LS is the long-short portfolio, formed by buying Q5 and selling Q1. Value-weighted portfolios use market capitalization as of the previous month. The returns are calculated at the monthly frequency over 1963-2018. I calculate $\beta_{CID}$ from abnormal returns of FF49 industries with respect to FF3 model.}
    \medskip
    \end{flushleft}
\begin{tabular}{@{\extracolsep{5pt}} ccccccc} 
\\[-1.8ex]\hline 
\hline \\[-1.8ex] 
 & Q1 & Q2 & Q3 & Q4 & Q5 & LS \\ 
\hline \\[-1.8ex] 
Mean ew & 0.76$^{***}$ & 0.79$^{***}$ & 0.72$^{***}$ & 0.68$^{***}$ & 0.52$^{**}$ & -0.24$^{**}$ \\ 
T-stat ew & [3.45] & [4.20] & [3.96] & [3.51] & [2.15] & [-2.17] \\ 
Mean vw & 0.78$^{***}$ & 0.63$^{***}$ & 0.51$^{***}$ & 0.53$^{***}$ & 0.31 & -0.48$^{***}$ \\ 
T-stat vw & [3.94] & [3.70] & [3.15] & [3.10] & [1.37] & [-3.20] \\ 
\hline \\[-1.8ex] 
\end{tabular}
\end{table}

\begin{table}[!htbp] \centering 
  \caption{\textbf{Abnormal returns of the long-short portfolio, formed on $\beta_{CID}$, calculated from abnormal returns of FF49 industry portfolios}} 
  \label{} 
    \begin{flushleft}
    {\medskip\small
 The table reports abnormal monthly returns of the long-short value-weighted quintile portfolio, formed from sorts on $\beta_{CID}$. The last column contains the abnormal returns with respect to Fama-French 5 factor model, augmented with momentum and short-term reversal factors. The returns are calculated at the monthly frequency over 1963-2018. I calculate $\beta_{CID}$ from abnormal returns of FF49 industries with respect to FF3 model.}
    \medskip
    \end{flushleft}
\begin{tabular}{@{\extracolsep{5pt}} ccccccc} 
\\[-1.8ex]\hline 
\hline \\[-1.8ex] 
Statistic & Ret & $\alpha_{CAPM}$ & $\alpha_{FF3}$ & $\alpha_{Carhart}$ & $\alpha_{FF5}$ & $\alpha_{FF5+UMD+STR}$ \\ 
\hline \\[-1.8ex] 
LS & -0.48$^{***}$ & -0.55$^{***}$ & -0.44$^{***}$ & -0.57$^{***}$ & -0.22 & -0.30$^{**}$ \\ 
 & [-3.20] & [-3.69] & [-2.97] & [-3.84] & [-1.47] & [-2.04] \\ 
\hline \\[-1.8ex] 
\end{tabular}
\end{table}

\begin{table}[!htbp] \centering 
  \caption{\textbf{Fama-MacBeth regression}} 
  \label{} 
  \begin{flushleft}
    {\medskip\small
 The table reports the results of Fama-MacBeth regression of excess returns on characteristics of the stocks. Size is log(ME) in the previous month. Momentum is the return over the past year, excluding the most recent month. Investment is defined as the growth in total assets over the recent year. MAX stands for the largest daily return over the previous month. All independent variables are winsorized at 10\% and 90\%. The returns are calculated at the monthly frequency over 1963-2018.}
    \medskip
    \end{flushleft}
\begin{tabular}{@{\extracolsep{0pt}}lccccccc} 
\\[-1.8ex]\hline 
\hline \\[-1.8ex] 
 & \multicolumn{7}{c}{\textit{Dependent variable: Return}} \\ 
\cline{2-8} 
\\[-1.8ex] & (1) & (2) & (3) & (4) & (5) & (6) & (7)\\ 
\hline \\[-1.8ex] 
 $\beta_{CID}$ & $-$0.05$^{**}$ & $-$0.03$^{*}$ & $-$0.02$^{*}$ & $-$0.02$^{*}$ & $-$0.02$^{**}$ & $-$0.02$^{**}$ & $-$0.02$^{**}$ \\ 
  & [$-$2.29] & [$-$1.87] & [$-$1.80] & [$-$1.72] & [$-$2.02] & [$-$2.00] & [$-$2.09] \\ 
  & & & & & & & \\ 
 $\beta$ &  & $-$0.21 & $-$0.20 & $-$0.07 & $-$0.29 & $-$0.23 & 0.13 \\ 
  &  & [$-$0.75] & [$-$0.63] & [$-$0.25] & [$-$1.00] & [$-$0.78] & [0.46] \\ 
  & & & & & & & \\ 
 size &  &  & $-$0.05 & $-$0.03 & $-$0.02 & $-$0.03 & $-$0.09$^{**}$ \\ 
  &  &  & [$-$1.15] & [$-$0.71] & [$-$0.51] & [$-$0.68] & [$-$2.38] \\ 
  & & & & & & & \\ 
 logbm &  &  &  & 0.22$^{***}$ & 0.22$^{***}$ & 0.15$^{***}$ & 0.11$^{**}$ \\ 
  &  &  &  & [3.83] & [3.93] & [2.68] & [2.12] \\ 
  & & & & & & & \\ 
 mom122 &  &  &  &  & 1.33$^{***}$ & 1.26$^{***}$ & 1.13$^{***}$ \\ 
  &  &  &  &  & [7.08] & [6.63] & [6.06] \\ 
  & & & & & & & \\ 
 inv &  &  &  &  &  & $-$0.94$^{***}$ & $-$0.93$^{***}$ \\ 
  &  &  &  &  &  & [$-$6.79] & [$-$6.76] \\ 
  & & & & & & & \\ 
 MAX &  &  &  &  &  &  & $-$0.08$^{***}$ \\ 
  &  &  &  &  &  &  & [$-$12.34] \\ 
  & & & & & & & \\ 
 Constant & 0.66$^{***}$ & 0.84$^{***}$ & 1.05$^{***}$ & 2.44$^{***}$ & 2.44$^{***}$ & 2.01$^{***}$ & 2.20$^{***}$ \\ 
  & [3.49] & [4.80] & [4.30] & [5.51] & [5.57] & [4.63] & [5.05] \\ 
  & & & & & & & \\ 
\hline \\[-1.8ex] 
Observations & 653 & 653 & 653 & 653 & 653 & 653 & 653 \\ 
R$^{2}$ & 0.0001 & 0.0003 & 0.0003 & 0.001 & 0.001 & 0.001 & 0.002 \\ 
Adjusted R$^{2}$ & 0.0001 & 0.0003 & 0.0003 & 0.001 & 0.001 & 0.001 & 0.002 \\ 
\hline 
\hline \\[-1.8ex] 
\textit{Note:}  & \multicolumn{7}{r}{$^{*}$p$<$0.1; $^{**}$p$<$0.05; $^{***}$p$<$0.01} \\ 
\end{tabular}
\end{table}

\clearpage

\begin{table}[!htbp] \centering 
  \caption{\textbf{Average returns of quintile $\beta_{CID}$-sorted portfolios, formed from 49 industry portfolios}} 
  \label{} 
  \begin{flushleft}
    {\medskip\small
 The table reports mean monthly excess returns of quintile portfolios, sorted on $\beta_{CID}$ from original 49 industry portfolios. Q1 is the quintile portfolio of industry portfolios with the lowest $\beta_{CID}$ and Q5 contains the highest $\beta_{CID}$ industry portfolios. L/S is the long-short portfolio, formed by buying Q5 and selling Q1. Value-weighted portfolios are constructed using as weights the market capitalization from the previous month. The returns are calculated at the monthly frequency over 1963-2018.}
    \medskip
    \end{flushleft}
\begin{tabular}{@{\extracolsep{5pt}} ccccccc} 
\\[-1.8ex]\hline 
\hline \\[-1.8ex] 
 & Q1 & Q2 & Q3 & Q4 & Q5 & L/S \\ 
\hline \\[-1.8ex] 
Mean ew & 0.98$^{***}$ & 0.91$^{***}$ & 0.96$^{***}$ & 0.92$^{***}$ & 0.84$^{***}$ & -0.14 \\ 
T-stat ew & [4.87] & [4.79] & [5.12] & [4.74] & [4.07] & [-1.06] \\ 
Mean vw & 1.09$^{***}$ & 0.86$^{***}$ & 1.04$^{***}$ & 0.85$^{***}$ & 0.78$^{***}$ & -0.31$^{*}$ \\ 
T-stat vw & [5.62] & [4.65] & [5.65] & [4.54] & [3.78] & [-1.90] \\ 
\hline \\[-1.8ex] 
\end{tabular} 
\end{table}

\begin{table}[!htbp] \centering 
  \caption{\textbf{Abnormal returns of quintile $\beta_{CID}$-sorted portfolios, formed from 49 industry portfolios}} 
  \label{} 
  \begin{flushleft}
    {\medskip\small
 The table reports abnormal monthly returns of the long-short value-weighted quintile portfolio, formed from sorts on $\beta_{CID}$ across original 49 industry portfolios. The last column contains the abnormal returns with respect to Fama-French 5 factor model, augmented with momentum and short-term reversal factors. The returns are calculated at the monthly frequency over 1963-2018.}
    \medskip
    \end{flushleft}
\begin{tabular}{@{\extracolsep{0pt}} ccccccc} 
\\[-1.8ex]\hline 
\hline \\[-1.8ex] 
Statistic & Ret & $\alpha_{CAPM}$ & $\alpha_{FF3}$ & $\alpha_{Carhart}$ & $\alpha_{FF5}$ & $\alpha_{FF5+UMD+STR}$ \\ 
\hline \\[-1.8ex] 
L/S & -0.31$^{*}$ & -0.36$^{**}$ & -0.20 & -0.43$^{***}$ & -0.09 & -0.31$^{**}$ \\ 
 & [-1.92] & [-2.21] & [-1.32] & [-2.82] & [-0.57] & [-1.98] \\ 
\hline \\[-1.8ex] 
\end{tabular} 
\end{table}

\begin{table}[!htbp] \centering 
  \caption{Predictive regressions for unemployment} 
  \label{} 
\begin{tabular}{@{\extracolsep{5pt}}lccc} 
\\[-1.8ex]\hline 
\hline \\[-1.8ex] 
 & \multicolumn{3}{c}{\textit{Dependent variable: Unemployment growth}} \\ 
\cline{2-4} 
\\[-1.8ex] & (1) & (2) & (3)\\ 
\hline \\[-1.8ex] 
 CIV & 4.02$^{***}$ & 3.97$^{***}$ & 3.69$^{***}$ \\ 
  & [2.77] & [2.90] & [2.85] \\ 
  & & & \\ 
 Mkt &  & $-$0.80$^{**}$ & $-$0.71$^{**}$ \\ 
  &  & [$-$2.53] & [$-$2.34] \\ 
  & & & \\ 
 Vol &  & 0.17$^{**}$ & 0.17$^{**}$ \\ 
  &  & [2.37] & [2.51] \\ 
  & & & \\ 
 FU &  &  & 0.83 \\ 
  &  &  & [0.60] \\ 
  & & & \\ 
 CIV &  &  & 0.02 \\ 
  &  &  & [0.37] \\ 
  & & & \\ 
 Constant & 0.005 & $-$0.01 & $-$0.01 \\ 
  & [0.10] & [$-$0.27] & [$-$0.27] \\ 
  & & & \\ 
\hline \\[-1.8ex] 
Observations & 284 & 220 & 220 \\ 
R$^{2}$ & 0.03 & 0.19 & 0.19 \\ 
Adjusted R$^{2}$ & 0.03 & 0.18 & 0.17 \\ 
\hline 
\hline \\[-1.8ex] 
\end{tabular} 
\end{table}

\begin{table}[!htbp] \centering 
  \caption{Predictive regressions for long-term unemployment} 
  \label{} 
\begin{tabular}{@{\extracolsep{5pt}}lccc} 
\\[-1.8ex]\hline 
\hline \\[-1.8ex] 
 & \multicolumn{3}{c}{\textit{Dependent variable: Unemployment growth}} \\ 
\cline{2-4} 
\\[-1.8ex] & (1) & (2) & (3)\\ 
\hline \\[-1.8ex] 
 CID & 2.65$^{**}$ & 2.88$^{***}$ & 3.22$^{***}$ \\ 
  & [2.15] & [2.62] & [3.14] \\ 
  & & & \\ 
 Mkt &  & $-$0.03 & $-$0.12 \\ 
  &  & [$-$0.14] & [$-$0.56] \\ 
  & & & \\ 
 Vol &  & 0.10$^{**}$ & 0.11$^{**}$ \\ 
  &  & [2.16] & [2.34] \\ 
  & & & \\ 
 FU &  &  & $-$0.21 \\ 
  &  &  & [$-$0.25] \\ 
  & & & \\ 
 CIV &  &  & $-$0.04 \\ 
  &  &  & [$-$0.91] \\ 
  & & & \\ 
 Constant & 0.01 & $-$0.002 & $-$0.002 \\ 
  & [0.19] & [$-$0.07] & [$-$0.08] \\ 
  & & & \\ 
\hline \\[-1.8ex] 
Observations & 284 & 220 & 220 \\ 
R$^{2}$ & 0.05 & 0.14 & 0.14 \\ 
Adjusted R$^{2}$ & 0.04 & 0.13 & 0.13 \\ 
\hline 
\hline \\[-1.8ex] 
\textit{Note:}  & \multicolumn{3}{r}{$^{*}$p$<$0.1; $^{**}$p$<$0.05; $^{***}$p$<$0.01} \\ 
\end{tabular} 
\end{table}

\begin{table}[!htbp] \centering 
  \caption{Predictive regressions for short-term unemployment} 
  \label{} 
\begin{tabular}{@{\extracolsep{5pt}}lccc} 
\\[-1.8ex]\hline 
\hline \\[-1.8ex] 
 & \multicolumn{3}{c}{\textit{Dependent variable:}} \\ 
\cline{2-4} 
\\[-1.8ex] & (1) & (2) & (3)\\ 
\hline \\[-1.8ex] 
 CID & 1.22$^{*}$ & 0.90 & 0.14 \\ 
  & [1.79] & [1.59] & [0.21] \\ 
  & & & \\ 
 Mkt &  & $-$0.80$^{***}$ & $-$0.57$^{***}$ \\ 
  &  & [$-$4.11] & [$-$3.49] \\ 
  & & & \\ 
 Vol &  & 0.07$^{*}$ & 0.06 \\ 
  &  & [1.88] & [1.64] \\ 
  & & & \\ 
 FU &  &  & 1.25 \\ 
  &  &  & [1.56] \\ 
  & & & \\ 
 CIV &  &  & 0.07$^{*}$ \\ 
  &  &  & [1.92] \\ 
  & & & \\ 
 Constant & $-$0.002 & $-$0.01 & $-$0.01 \\ 
  & [$-$0.08] & [$-$0.49] & [$-$0.45] \\ 
  & & & \\ 
\hline \\[-1.8ex] 
Observations & 286 & 220 & 220 \\ 
R$^{2}$ & 0.01 & 0.18 & 0.21 \\ 
Adjusted R$^{2}$ & 0.004 & 0.17 & 0.19 \\ 
\hline 
\hline \\[-1.8ex] 
\textit{Note:}  & \multicolumn{3}{r}{$^{*}$p$<$0.1; $^{**}$p$<$0.05; $^{***}$p$<$0.01} \\ 
\end{tabular} 
\end{table}

\begin{table}[!htbp] \centering 
  \caption{Returns of portfolios, double-sorted on $\beta_{CID}$ and $\beta_{NVIX}$} 
  \label{} 
    \begin{flushleft}
    {\medskip\small
 The table reports monthly returns of long-short value-weighted quintile portfolios, formed from independent sorts on $\beta_{CID}$ and $\beta_{NVIX}$. NVIX is news-implied volatility (Manela and Moreira, 2017.}
    \medskip
    \end{flushleft}
\begin{tabular}{@{\extracolsep{5pt}} ccccccc} 
\\[-1.8ex]\hline 
\hline \\[-1.8ex] 
Statistic & Ret & $\alpha_{CAPM}$ & $\alpha_{FF3}$ & $\alpha_{Carhart}$ & $\alpha_{FF5}$ & $\alpha_{FF5+UMD+STR}$ \\ 
\hline \\[-1.8ex] 
L/S NVIX & 0.04 & -0.10 & -0.03 & 0.06 & 0.25$^{*}$ & 0.36$^{***}$ \\ 
T-stat & [ 0.26] & [ -0.72] & [ -0.19] & [ 0.42] & [ 1.88] & [ 2.63] \\ 
L/S CID & 0.45$^{***}$ & 0.49$^{***}$ & 0.33$^{**}$ & 0.55$^{***}$ & 0.23 & 0.43$^{***}$ \\ 
T-stat & [ 2.99] & [ 3.26] & [ 2.28] & [ 3.82] & [ 1.54] & [ 2.94] \\ 
\hline \\[-1.8ex] 
\end{tabular} 
\end{table}

\newpage

\begin{table}[!htbp] \centering 
  \caption{Quintile returns using quarterly $\beta_{CID}$} 
  \label{} 
    \begin{flushleft}
    {\medskip\small
 The table reports mean monthly excess returns of quintile portfolios, sorted on $\beta_{CID}$. Q1 is the quintile portfolio with the lowest $\beta_{CID}$. The returns are calculated at the monthly frequency over 1963-2019.}
    \medskip
    \end{flushleft}
\begin{tabular}{@{\extracolsep{5pt}} ccccccc} 
\\[-1.8ex]\hline 
\hline \\[-1.8ex] 
 & Q1 & Q2 & Q3 & Q4 & Q5 & LS \\ 
\hline \\[-1.8ex] 
Mean ew & 0.82$^{***}$ & 0.77$^{***}$ & 0.76$^{***}$ & 0.73$^{***}$ & 0.63$^{***}$ & -0.19$^{*}$ \\ 
T-stat ew & [4.23] & [4.43] & [4.24] & [3.91] & [2.79] & [-1.90] \\ 
Mean vw & 0.66$^{***}$ & 0.57$^{***}$ & 0.51$^{***}$ & 0.48$^{***}$ & 0.45$^{**}$ & -0.22 \\ 
T-stat vw & [3.57] & [3.59] & [3.07] & [2.74] & [2.21] & [-1.51] \\ 
\hline \\[-1.8ex] 
\end{tabular} 
\end{table}

\begin{table}[!htbp] \centering 
  \caption{Quintile abnormal returns using quarterly $\beta_{CID}$} 
  \label{} 
    \begin{flushleft}
    {\medskip\small
 The table reports abnormal monthly returns of the long-short value-weighted quintile portfolio, formed from sorts on $\beta_{CID}$. The last column contains the abnormal returns with respect to Fama-French 5 factor model, augmented with momentum and short-term reversal factors. The returns are calculated at the monthly frequency over 1963-2019.}
    \medskip
    \end{flushleft}
\begin{tabular}{@{\extracolsep{5pt}} ccccccc} 
\\[-1.8ex]\hline 
\hline \\[-1.8ex] 
Statistic & Ret & $\alpha_{CAPM}$ & $\alpha_{FF3}$ & $\alpha_{Carhart}$ & $\alpha_{FF5}$ & $\alpha_{FF5+UMD+STR}$ \\ 
\hline \\[-1.8ex] 
LS & -0.22 & -0.29$^{**}$ & -0.19 & -0.38$^{***}$ & -0.24$^{*}$ & -0.40$^{***}$ \\ 
 & [-1.51] & [-2.06] & [-1.38] & [-2.86] & [-1.69] & [-2.90] \\ 
\hline \\[-1.8ex] 
\end{tabular} 
\end{table}

\begin{table}[!htbp] \centering 
  \caption{Quintile return using daily $\beta_{CID}$}
  \label{} 
      \begin{flushleft}
    {\medskip\small
 The table reports mean monthly excess returns of quintile portfolios, sorted on $\beta_{CID}$. Q1 is the quintile portfolio with the lowest $\beta_{CID}$. The returns are calculated at the monthly frequency over 1963-2019.}
    \medskip
    \end{flushleft}
\begin{tabular}{@{\extracolsep{5pt}} ccccccc} 
\\[-1.8ex]\hline 
\hline \\[-1.8ex] 
 & Q1 & Q2 & Q3 & Q4 & Q5 & LS \\ 
\hline \\[-1.8ex] 
Mean ew & 0.70$^{***}$ & 0.79$^{***}$ & 0.78$^{***}$ & 0.66$^{***}$ & 0.54$^{**}$ & -0.17 \\ 
T-stat ew & [3.15] & [4.20] & [4.13] & [3.33] & [2.27] & [-1.56] \\ 
Mean vw & 0.69$^{***}$ & 0.65$^{***}$ & 0.58$^{***}$ & 0.54$^{***}$ & 0.38$^{*}$ & -0.31$^{**}$ \\ 
T-stat vw & [3.43] & [3.86] & [3.55] & [3.29] & [1.92] & [-2.29] \\ 
\hline \\[-1.8ex] 
\end{tabular} 
\end{table}

\begin{table}[!htbp] \centering 
  \caption{Quintile abnormal returns using daily $\beta_{CID}$} 
  \label{} 
    \begin{flushleft}
    {\medskip\small
 The table reports abnormal monthly returns of the long-short value-weighted quintile portfolio, formed from sorts on $\beta_{CID}$. The last column contains the abnormal returns with respect to Fama-French 5 factor model, augmented with momentum and short-term reversal factors. The returns are calculated at the monthly frequency over 1963-2019.}
    \medskip
    \end{flushleft}
\begin{tabular}{@{\extracolsep{5pt}} ccccccc} 
\\[-1.8ex]\hline 
\hline \\[-1.8ex] 
Statistic & Ret & $\alpha_{CAPM}$ & $\alpha_{FF3}$ & $\alpha_{Carhart}$ & $\alpha_{FF5}$ & $\alpha_{FF5+UMD+STR}$ \\ 
\hline \\[-1.8ex] 
LS & -0.31$^{**}$ & -0.31$^{**}$ & -0.35$^{***}$ & -0.19 & -0.34$^{***}$ & -0.13 \\ 
 & [-2.29] & [-2.34] & [-2.75] & [-1.51] & [-2.61] & [-0.96] \\ 
\hline \\[-1.8ex] 
\end{tabular} 
\end{table}

\begin{table}[!htbp] \centering 
  \caption{Correlations between L/S portfolios, formed on $\beta_{CID}$ at varying frequencies} 
  \label{} 
  \begin{flushleft}
    {\medskip\small
 The table reports abnormal monthly returns of the long-short value-weighted quintile portfolio, formed from sorts on $\beta_{CID}$. The last column contains the abnormal returns with respect to Fama-French 5 factor model, augmented with momentum and short-term reversal factors. The returns are calculated at the monthly frequency over 1963-2019.}
    \medskip
    \end{flushleft}
\begin{tabular}{@{\extracolsep{5pt}} ccc} 
\\[-1.8ex]\hline 
\hline \\[-1.8ex] 
Quarterly\_LS & Monthly\_LS & Daily\_LS \\ 
\hline \\[-1.8ex] 
$1$ & $0.26$ & $-0.04$ \\ 
$0.26$ & $1$ & $0.06$ \\ 
$-0.04$ & $0.06$ & $1$ \\ 
\hline \\[-1.8ex] 
\end{tabular} 
\end{table}

\begin{table}[!htbp] \centering 
  \caption{Predictbility of industry employment by industry returns at quarterly frequency} 
  \label{}
    \begin{flushleft}
    {\medskip\small
 The table reports results of predictive panel regressions employment growth of 14 industries using their lagged returns. Employment growths and returns are normalized by aggregate employmnet growth and market return. The variables are calculated at the quaterly frequency over 1990-2019.}
    \medskip
    \end{flushleft}
\begin{tabular}{@{\extracolsep{5pt}}lccc} 
\\[-1.8ex]\hline 
\hline \\[-1.8ex] 
 & \multicolumn{3}{c}{\textit{Dependent variable: Employment growth}} \\ 
\cline{2-4} 
\\[-1.8ex] \\
\hline \\[-1.8ex] 
  $Return_{t-1}$ & 0.013$^{***}$ & 0.020$^{***}$ & 0.007$^{*}$ \\ 
  & (0.002) & (0.006) & (0.004) \\ 
  & & & \\ 
 Constant & $-$0.107$^{***}$ & $-$0.073$^{*}$ & $-$0.070$^{**}$ \\ 
  & (0.018) & (0.041) & (0.034) \\ 
  & & & \\ 
\hline \\[-1.8ex] 
Sample & Full & Negative $R_{t-1}$ & Positive $R_{t-1}$ \\
Observations & 1,680 & 842 & 838 \\ 
Adjusted R$^{2}$ & 0.018 & 0.014 & 0.002 \\ 
\hline 
\hline \\[-1.8ex] 
\textit{Note:}  & \multicolumn{3}{r}{$^{*}$p$<$0.1; $^{**}$p$<$0.05; $^{***}$p$<$0.01} \\ 
\end{tabular} 
\end{table}

\end{document}